\documentclass[pre,aps,twocolumn,superscriptaddress]{revtex4}
\usepackage[dvips]{graphicx}
\usepackage{amssymb,amsfonts,amsmath}
\usepackage{xcolor}
\usepackage{ulem}
\usepackage[hidelinks]{hyperref}

\newcommand{\rv}{{\mathbf r}}
\newcommand{\ov}{{\boldsymbol{\omega}}}
\newcommand{\ev}{{\bf e}}

\newcommand{\Jv}{{\bf J}}

\newcommand{\Fv}{{\bf F}}
\newcommand{\fv}{{\bf f}}

\newcommand{\vel}{{\bf v}}

\newcommand{\chib}{{\boldsymbol \chi}}

\newcommand{\ms}{\color{black}}

\newcommand{\mz}{\color{black}}
\newcommand{\mm}{\color{black}}
\newcommand{\mb}{\color{black}}
\newcommand{\mc}{\color{black}}
\begin{document}

\title{Phase coexistence of active Brownian particles}

\author{Sophie Hermann}
\affiliation{Theoretische Physik II, Physikalisches Institut, 
  Universit{\"a}t Bayreuth, D-95447 Bayreuth, Germany}

\author{Philip Krinninger}
\affiliation{Theoretische Physik II, Physikalisches Institut, 
  Universit{\"a}t Bayreuth, D-95447 Bayreuth, Germany}

\author{Daniel de las Heras}
\affiliation{Theoretische Physik II, Physikalisches Institut, 
  Universit{\"a}t Bayreuth, D-95447 Bayreuth, Germany}

\author{Matthias Schmidt}
\affiliation{Theoretische Physik II, Physikalisches Institut, 
  Universit{\"a}t Bayreuth, D-95447 Bayreuth, Germany}
\email{Matthias.Schmidt@uni-bayreuth.de}

\date{28 December 2018, \href{https://link.aps.org/doi/10.1103/PhysRevE.100.052604}{Phys. Rev. E \textbf{100}, 052604 (2019).}}

\begin{abstract} 
We investigate motility-induced phase separation of active Brownian
particles, which are modeled as purely repulsive spheres that move due
to a constant swim force with freely diffusing orientation. We develop
{\ms on the basis of power functional concepts} an analytical theory
for nonequilibrium phase coexistence and interfacial
structure. Theoretical predictions are validated against Brownian
dynamics computer simulations.  We show that the internal one-body
force field has four nonequilibrium contributions: (i) isotropic drag
and (ii) interfacial drag forces against the forward motion, (iii) a
superadiabatic spherical pressure gradient and (iv) the quiet life
gradient force.  The intrinsic spherical pressure is balanced by the
swim pressure, which arises from the polarization of the free
interface.  The quiet life force opposes the adiabatic force, which is
due to the inhomogeneous density distribution. The balance of quiet
life and adiabatic forces determines bulk coexistence via equality of
two bulk state functions, which are independent of interfacial
contributions. The internal force fields are kinematic functionals
which depend on density and current, but are independent of external
and swim forces, consistent with power functional theory. The phase
transition originates from nonequilibrium repulsion, with the agile
gas being more repulsive than the quiet liquid.
\end{abstract}

\maketitle 

\section{Introduction}
The spontaneous occurrence of gas-liquid phase separation into
macroscopic bulk phases and the associated emergence of a stable
interface between the two different fluids is one of the most striking
phenomena in equilibrium statistical physics. It was Johannes van der
Waals who first developed microscopic theories for both the bulk
behaviour \cite{vanderWaals1873} and the interfacial structure
\cite{vanderWaals1893}. Subsequently, Smoluchowski
\cite{smoluchowski1908} and Mandelstam \cite{mandelstam1913}
successfully described thermally excited capillary waves as collective
fluctuations that generate interfacial roughness. The description of
the free fluid interface constitutes one of the most challenging
problems in statistical mechanics
\cite{RowlinsonWidomBook,Hansen06,evans1979}. It is relevant for the
demixing of liquid mixtures \cite{zhang2018,ashton2011}, and it forms
one of the most well-developed cornerstones of theoretical physics
\cite{triezenberg72,weeks77,wertheim,parry16}. Both advanced computer
simulations \cite{chacon03prlIntrinsic} and direct experimental
observation, in e.g.\ colloid-polymer mixture \cite{aarts2004}, are
means of investigation.

Given this situation in equilibrium, it seems natural to attempt to
describe fluid interfaces in nonequilibrium on the basis of similar
concepts.  Examples of this strategy include the reduction of
experimentally observed \cite{derks2008} interfacial roughness via
shear flow as an effective confinement effect \cite{smith2008}. In the
context of active fluids, which consist of self-driven particles,
integrating out the swimming was shown to lead to an effective
attraction between the particles \cite{farage2015effective}, which
then can be input into an equilibrium treatment of the interfacial
structure, e.g.\ on the basis of classical density functional theory
\cite{wittman2016interface}.

Active Brownian particles have become a prototype for the study of
nonequilibrium phenomena
\cite{maggi2015,marconi2015towards}. In particular their
``motility-induced'' phase separation into high- and low-density
steady states continues to attract much current interest
\cite{paliwal2018njp,digregorio2018prl,solon2018rapComm,solon2018njp,
  prymidis16activeLJ,vanderMeer16activepassive}. Very significant
efforts have been devoted to understanding this phase transition,
which occurs without any explicit interparticle attraction. This is a
striking difference to the equilibrium gas-liquid case, where the
balance of short-ranged repulsion and long-ranged intermolecular
attraction drives a transition between the gas with high entropy and
high energy and the liquid with low entropy and low energy. In
contrast, for the motility-induced case, frequently a ``feedback''
mechanism is invoked in which particles in dense regions slow down
\cite{feedback}. The striking feature of the transition is the very
strong inhomogeneity in density between the dense and the dilute
phase. The challenge lies in understanding what physical mechanism
would oppose the strong tendency of the liquid to expand and hence to
homogenize the system. The homogenization does not occur, at strong
enough driving conditions, such that nonequilibrium phase coexistence
is stable.

That strong density inhomogeneities can spontaneously occur is
well-known in equilibrium situations. Examples include adsorption of
liquid or solid films on substrates and capillary condensation and
freezing inside of narrow pores as well as nucleation phenomena
\cite{lutsko}. In these equilibrium cases balancing forces have been
identified that act against the interparticle repulsion, be it
intermolecular attraction, such as in the Lennard-Jones system, or via
the influence of further species that generate effective attraction
via the depletion mechanism. There are prominent cases, such as
e.g.\ colloid-polymer mixtures, where fluid-fluid phase separation
occurs in purely repulsive mixtures. In monocomponent system, the hard
sphere fluid-solid (freezing) transition is a further prototype for
coexistence of phases with differing spatial symmetries.

Often the interparticle repulsion is satisfactorily treated in a local
way, assuming that high local density is associated with a free energy
penalty that is taken to be a function of the local
density. Nevertheless, much more sophisticated approximations exist
within the framework of classical density functional theory, where a
broad range of approximate functionals is available, ranging from
square-gradient semi-local functionals to fully-nonlocal
fundamental-measure Rosenfeld theories.  When the system is driven out
of equilibrium, then additional force contributions arise. In
nonequilibrium, construction of the adiabatic state
\cite{fortini14prl} allows to systematically rationalize the force
field that is solely due to the inhomogeneous density distribution,
and subsequently analyse systematically the additional nonequilibrium
(superadiabatic) forces.

Active phase separation is considered to be such a very striking
phenomenon, as no apparent balancing mechanism, which would counteract
the repulsion and keep the dense region compressed, has been
identified.  Nevertheless, a broad variety of different theoretical
methods have been employed to study the phase separation phenomenon,
which occurs very prominently and in a robust and reproducible way in
Brownian dynamics (BD) computer simulations
\cite{digregorio2018prl,paliwal2018njp,solon2018rapComm,solon2018njp,prymidis16activeLJ,vanderMeer16activepassive}. Among
the different theoretical approaches are theories based on modified
forms of the Cahn-Hilliard equation
\cite{solon2018rapComm,solon2018njp,speck2014prl}, hydrodynamic
description \cite{stark2016sm}, and more microscopic statistical
mechanics treatments that start from the Smoluchowski equation of
motion for the many-body probability distribution function
\cite{paliwal2018njp}. One closely related aim is to identify
coexistence conditions and to construct a thermodynamic description of
the system \cite{takatoriBrady2015thermodynamics}. This is relevant as
it allows to judge whether and if so which of the properties of
equilibrium gas-liquid phase separation carry over to the
nonequilibrium case.
The most recent treatments conclude that interfacial effects affect
the bulk coexistence
\cite{paliwal2018njp,solon2018rapComm,solon2018njp}, in striking
contrast to the equilibrium case.

Here we study the bulk behaviour and interfacial structure of active
Brownian particles in nonequilibrium steady states. We develop an
analytical theory for the free interface between phase-separated bulk
states of active Brownian particles. The theory is fully resolved in
both position and orientation.  The symmetry of the problem allows to
reduce the dependence on one spatial coordinate ($x$) perpendicular to
the interface and one angle ($\varphi$) of particle orientation
against the $x$-axis. The theory describes correctly the orientational
ordering at the interface, including dipolar and higher orientational
moments. We validate the theoretical results for the force fields
against (overdamped) BD simulation data of the phase separated
system. Bulk phase coexistence occurs on the isotropic level of the
correlation functions and the coexistence conditions are independent
of interfacial effects.  As an illustration, we show a BD simulation
snapshot in Fig.~\ref{fig1}(a), obtained for the frequently-used
Weeks-Chandler-Anderson (WCA) repulsive pair potential $\phi(r)$
\cite{WCA1971}, as plotted in Fig.~\ref{fig1}(b).

\section{Theory}
{\ms
\subsection{Many-body dynamics}
In the Langevin picture, the many-body
dynamics of $N$ active Brownian particles are given by
\begin{align}
  \gamma \dot\rv_i &= 
  -\nabla_i \sum_{j(\neq i)}\phi(|\rv_i-\rv_j|)
  +\gamma s\ov_i +\chib_i,
  \label{EQlangevinPositions}
\end{align}
where $\gamma$ is the friction constant against the static background,
$\rv_i(t)$ indicates the position of particle $i=1,\ldots,N$ at time
$t$, the overdot indicates a time derivative, $\nabla_i$ indicates the
derivative with respect to $\rv_i$, $s=\rm const$ is the speed of free
swimming (such that $\gamma s$ is the magnitude of the swim force),
the unit vector $\ov_i$ denotes the orientational degrees of freedom
(along which the swim force acts) of particle $i$, and $\chib_i(t)$ is
a stochastic white noise force term, which is bias free,
$\langle\chib_i(t)\rangle=0$, and delta correlated with itself,
$\langle \chib_i(t)\chib_j(t')\rangle=2k_BT\gamma \delta(t-t'){\bf
  1}\delta_{ij}$. Here the angles denote an average of the noise, $\bf
1$ denotes the $d\times d$ unit matrix, where $d$ is the space
dimensionality, $k_B$ denotes the Boltzmann constant, and~$T$
indicates absolute temperature; the translational diffusion constant
is then given by $D=k_BT/\gamma$.  {\mz In a system with $d=2$ space
  dimensions, as we consider below, the particle orientations can be
  parametrized by the angle $\varphi_i$ of particle orientation
  $\ov_i$ against the $x$-axis,
  i.e.\ $\ov_i=(\cos\varphi_i,\sin\varphi_i)$. The particle
  orientations diffuse freely, and hence
\begin{align}
  \gamma^\omega\dot\varphi_i &= \chi_i,
  \label{EQlangevinOrientations}
\end{align}
where $\chi_i(t)$ is an angular noise term with vanishing mean,
$\langle\chi_i(t)\rangle=0$, and auto-correlation given by
$\langle\chi_i(t)\chi_j(t')\rangle=2k_BT\gamma^\omega\delta(t-t')\delta_{ij}$.
(For a description of rotational diffusion in $d=3$ see,
e.g.\ \cite{farage2015effective}.) }
For completeness, the rotational diffusion constant
is then $D_{\rm rot}=k_BT/\gamma^\omega$.  The BD simulations are
based on the (standard) Euler algorithm for the system of equations
\eqref{EQlangevinPositions} and \eqref{EQlangevinOrientations} with
time discretization step $\Delta t$.
}

\subsection{Force density balance}
We operate on the level of position- and orientation-resolved one-body
fields: the one-body density $\rho(\rv,\ov,t)$, the translational
current $\Jv(\rv,\ov,t)$, and the rotational current
$\Jv^\omega(\rv,\ov,t)$, where $\rv$ denotes position, the unit vector
$\ov$ denotes the particle orientation along which the swimming force
acts, and~$t$ denotes time. The one-body fields are related by the
exact (translational) force density balance,
\begin{align}
  \gamma\Jv &= \gamma s \ov \rho + \Fv_{\rm int} - k_BT\nabla\rho,
  \label{EQforceDensityBalance}
\end{align}
where the arguments $\rv,\ov,t$ of the three one-body fields
$\Jv,\rho$ and $\Fv_{\rm int}$ have been omitted for clarity.  The
force density balance~\eqref{EQforceDensityBalance} expresses the
equality of the friction force density (left hand side) with the sum
of the driving that generates the swimming (first term on the right
hand side), the internal force density $\Fv_{\rm int}(\rv,\ov,t)$
(second term), and the thermal diffusion (third term); {\ms see
  appendix \ref{APPoneBodyFromSmoluchowski} for a derivation from the
  many-body dynamics.}  The rotational motion alone is simple: Due to
the free rotational diffusion (the particles are spherical), the
rotational current is simply $\Jv^\omega(\rv,\ov,t)=-D_{\rm rot}
\nabla^\omega\rho(\rv,\ov,t)$, where $D_{\rm rot}$ is the rotational
diffusion constant, and $\nabla^\omega$ is the derivative in the space
of orientations. The continuity equation is
\begin{align}
  \frac{\partial\rho}{\partial t} =
  - \nabla\cdot\Jv-\nabla^\omega \cdot \Jv^\omega,
  \label{EQcontinuity}
\end{align}
with $\partial \rho/\partial t=0$ in steady state.

The internal force density field, as occurring in
\eqref{EQforceDensityBalance}, was proven to be a ``kinematic''
functional of the density and the current
\cite{schmidt2013pft,delasheras2018customFlow}, i.e.\ $\Fv_{\rm
  int}(\rv,\ov,t)=\Fv_{\rm int}([\rho,\Jv,\Jv^\omega],\rv,\ov,t)$. No
further ``hidden'' dependence occurs; the internal force density field
is in particular independent of external and swim forces. This theorem
applies in general overdamped Brownian systems; see
\cite{krinninger2016prl,krinninger2019jcp} for the generalization of
power functional theory \cite{schmidt2013pft} to rotator models, such
as the current one.

\begin{figure*}
  \includegraphics[width=0.99\textwidth]{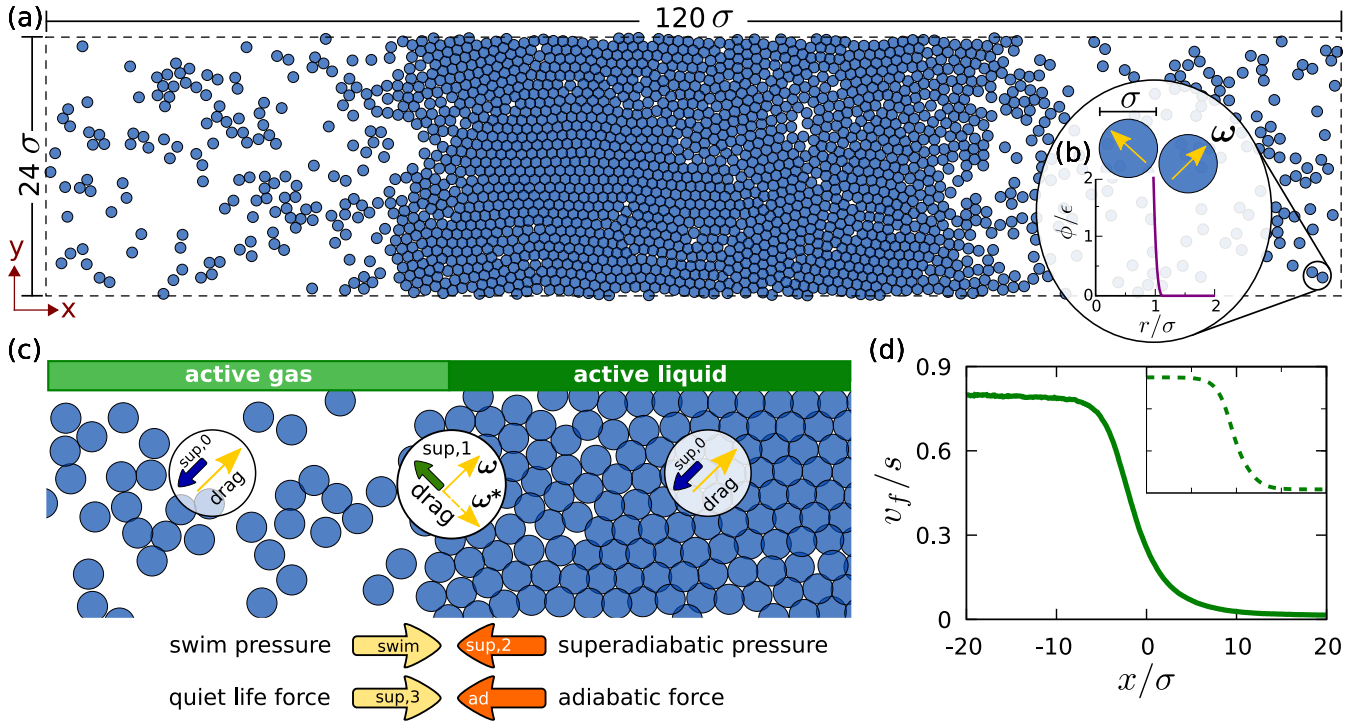}
  \caption{Overview of phase separated active Brownian particles.  (a)
    Snapshot from BD simulations. The particles (blue) separate into
    regions with high and with low density. Due to periodic boundary
    conditions in $x$ and in $y$, two interfaces form along the short
    dimension of the simulation box with size $120\sigma$ in $x$ and
    $24\sigma$ in $y$.  (b) Particles with size $\sigma$ at position
    $\rv=(x,y)$ are driven in (unit vector) direction $\ov$ by the
    swim force with strength $\gamma s$. The particles interact with
    the (WCA) repulsion $\phi(r)$ with energy scale $\epsilon$.  (c)
    Illustration of the different one-body force density contributions
    that add up to the total internal force density field $\Fv_{\rm
      int}$: the drag force density $\Fv_{\rm sup,0}$ acts against the
    local current direction $\Jv$ and its magnitude is small (large)
    in the dilute (dense) phase. The non-spherical drag correction
    $\Fv_{\rm sup,1}$ occurs at the interface and it acts in direction
    $\ov^*$ ($\ov$ mirrored at the $x$-axis). The swim pressure
    (orange arrow) is due to the polarization of the interface and it
    is balanced by the superadiabatic pressure (red arrow), which is
    low (high) in the dilute (dense) phase. The arrows indicate the
    direction of the respective negative pressure gradient.  The quiet
    life force field $\Fv_{\rm sup,3}/\rho$ compresses the liquid and
    acts against the adiabatic force field $\Fv_{\rm ad}/\rho$, which
    is (solely) due to the density gradient and tends to expand the
    liquid.  (d) Mean scaled forward swimming speed $v_{\rm f}/s$ as a
    function of the scaled position $x/\sigma$ across the interface in
    a phase-separated system of active Brownian particles interacting
    with the WCA pair potential, with particle size $\sigma$ and
    energy scale $\epsilon$. Simulation data are shown for
    $k_BT/\epsilon=0.5$ and $s{\mm \tau}/\sigma=60$, {\mm where the
      timescale is $\tau=\sigma^2\gamma/\epsilon$}, which corresponds
    to ${\rm Pe}=120$. The aspect ratio of the simulation box is 5 and
    the number $N$ of particles per system volume $V$ is
    $N/V=0.7\sigma^{-2}$ with $N=2000$; {\mm the time step is $\Delta
      t/\tau = 10^{-5}$. Sampling was performed over $10^8$ time
      steps; see \cite{krinninger2019jcp} for further simulation
      details that also apply to the present study.} The inset shows
    the theoretical result \eqref{EQvfAsFunctional}.
    {\mm The inset axis labels have been omitted for clarity; the
      scale is identical to that of the main plot. }}
\label{fig1}
\end{figure*}

\begin{figure}
  \includegraphics[width=0.98\columnwidth]{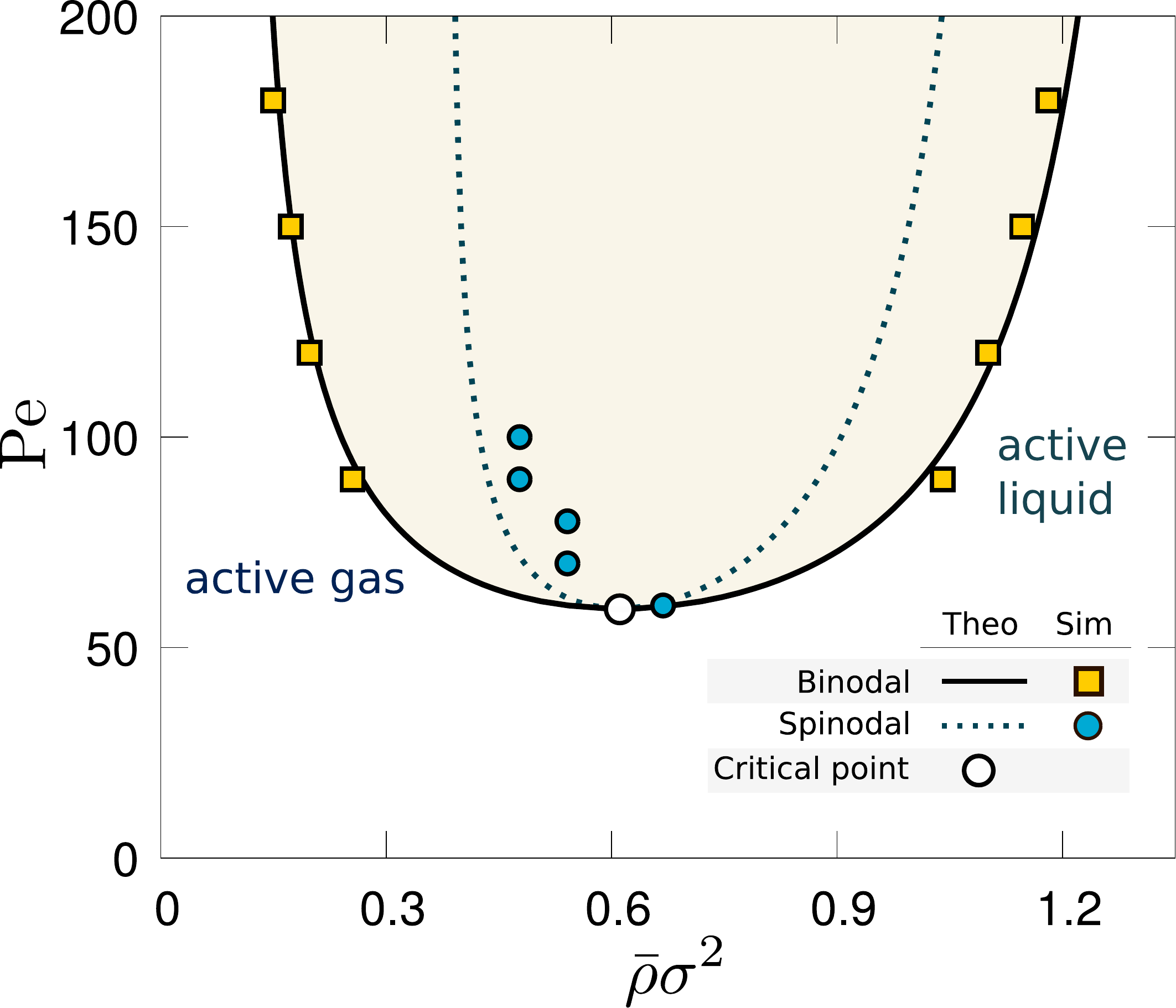}
  \caption{Nonequilibrium phase diagram as a function of scaled
    density $\rho_{\rm b}\sigma^2$ and Peclet number $\rm Pe$. Shown
    are the nonequilibrium binodal (solid line) and spinodal (dashed
    line) obtained from the present theory, compared to results for
    the binodal (orange squares) taken from
    Ref.~\cite{paliwal2018njp}, and for the gas side of the spinodal
    (blue circles) taken from Ref.~\cite{stenhammar2013}. Also shown
    is the theoretical result for the critical point (open circle).}
  \label{figPhaseDiagram}
\end{figure}

{\mz The internal force density field splits into a sum of adiabatic
  and superadiabatic contributions,
\begin{align}
  \Fv_{\rm int} &= \Fv_{\rm ad} + \Fv_{\rm sup},
  \label{EQforceDensitySplittingAdSup}
\end{align}
where $\Fv_{\rm ad}$ is the force density in a corresponding
``adiabatic'' system. The adiabatic system is in equilibrium and
constructed in such a way that its one-body density profile is
identical to that of the true nonequilibrium system. The Mermin-Evans
theorem of classical density functional theory \cite{evans1979}
ensures that an external potential exists in the adiabatic system that
accomplishes this task. We give a brief summary of classical density
functional theory in appendix \ref{SECdft}. Crucially $\Fv_{\rm ad}$
depends (functionally) only on the density profile and not on the
external force field.  {\mm Furthermore $\Fv_{\rm ad}$ is independent
  of the (translational and rotational) current.  The splitting
  \eqref{EQforceDensitySplittingAdSup} is exact; it is a consequence
  of the power functional variational framework
  \cite{schmidt2013pft,krinninger2019jcp}, and it was explicitly
  demonstrated in computer simulation work \cite{fortini14prl}.  In
  contrast to the adiabatic contribution, the superadiabatic force
  density profile $\Fv_{\rm sup}$ is a functional of the current (in
  the present case translational current and rotational current) as
  well as of the density profile.  }

The excess adiabatic force density field can be expressed as $\Fv_{\rm
  ad}=-\rho \nabla\delta F_{\rm exc}[\rho]/\delta\rho$, with the
excess (over ideal gas) Helmholtz free energy density functional
$F_{\rm exc}[\rho]$ \cite{evans1979}, {\mz see appendix \ref{SECdft}}.
For the present spherically repulsive interparticle interaction
potential, $\Fv_{\rm ad}$ describes the repulsion that is solely due
to the inhomogeneous density distribution.  The genuine nonequilibrium
contribution in \eqref{EQforceDensitySplittingAdSup} is the
superadiabatic force density profile $\Fv_{\rm sup}$. Power functional
theory \cite{schmidt2013pft,krinninger2019jcp} ensures that $\Fv_{\rm
  sup}$ depends both on the density profile and on the current
distribution, but not explicitly on the external force field.

We proceed by splitting the total superadiabatic force density
distribution into four different parts
\begin{align}
  \Fv_{\rm sup} &= \Fv_{\rm sup,0}
  +\Fv_{\rm sup,1}+\Fv_{\rm sup,2}+\Fv_{\rm sup,3},
  \label{EQFsupSplitting}
\end{align}
where the contributions $\Fv_{\rm sup,0}, \Fv_{\rm sup,1}$ and
$\Fv_{\rm sup,2}$ are obtained via projection of $\Fv_{\rm int}$ onto
a corresponding relevant orientation in the system and suitable
averaging. The remainder is then contained in $\Fv_{\rm sup,3}$, and
hence \eqref{EQFsupSplitting} does not constitute an approximation.
The mathematical structure of each superadiabatic term in
\eqref{EQFsupSplitting} is unique and characterizes a specific
physical effect. The projections are performed by the correlator
expressions for $\Fv_{\rm sup,0}$, $\Fv_{\rm sup,1}$, and $\Fv_{\rm
  sup,2}$, which are given and discussed below in
\eqref{EQFsup0PrimeAsCorrelator}, \eqref{EQFsup1PrimeAsAverage} and
\eqref{EQFsup2AsCorrelator}, respectively. Here we first briefly
discuss these individual terms before going into more detail.

The drag force density $\Fv_{\rm sup,0}$ acts against the local flow
direction. It leads to slowing of the forward swimming due to
collisions of the particle at position $\rv$ with surrounding
particles. The strength of this effect increases the more crowded the
environment is. The interfacial drag force density $\Fv_{\rm sup,1}$
is an orientation-dependent interfacial contribution with
``tensorial'' character, i.e.\ this drag force is not directed
strictly against the forward motion, but takes account of the gradient
direction in the system. This effect is induced by the inhomogeneous
environment at the interface.

The force density field $\Fv_{\rm sup,2}$ is the negative gradient of
an intrinsic spherical pressure $\Pi_2$. Characteristically $\Pi_2$ is
independent of orientation $\ov$. As a result the corresponding force
density $\Fv_{\rm sup,2}=-\nabla \Pi_2$ is also independent of
$\ov$. Similarly, the quiet life force field $\Fv_{\rm sup,3}/\rho$ is
the (negative) gradient of a spherical (nonequilibrium) chemical
potential $\nu_3$, i.e.\ $\Fv_{\rm sup,3}/\rho = -\nabla\nu_3$.  All
superadiabatic contributions describe repulsion and occur due to the
nonequilibrium driving in the system. We show below how these
repulsive forces generate stable phase coexistence.  As we
demonstrate, in particular the quiet life force stabilizes the
nonequilibrium bulk phase coexistence. It tends to push particles into
the liquid where the mean velocity is low (towards the ``quiet
life'').  Except for the bulk contribution to $\Fv_{\rm sup,0}$,
neither of the internal force density contributions has been
identified before. Figure \ref{fig1}(c) displays a schematic overview
of all forces that act in the system.}

The motion in the system is characterized by the forward current
profile $J_{\rm f}(x)$, defined as an angular average of the
projection of the current $\Jv$ onto the particle orientation $\ov$,
\begin{align}
  J_{\rm f} &= \frac{1}{2\pi}\int d\ov \Jv\cdot\ov.
  \label{EQJfAsAverage}
\end{align}
The corresponding forward swimming speed profile $v_{\rm f}(x)$ is
then obtained simply via
\begin{align}
  v_{\rm f} &= J_{\rm f}/\rho_0,
  \label{EQvfAsAverage}
\end{align}
where the angular average of the density profile is defined as
$\rho_0=\int d\ov \rho/(2\pi)$.  As an illustration, we show in
Fig.~\ref{fig1}(d) BD results for $v_{\rm f}$ as a function of $x$
across the interface in the phase separated system. The forward speed
is high in the gas and low in the liquid
\cite{fily2012prl,stenhammar2013} and it crosses over smoothly between
these plateau values as $x$ is varied from one phase to the other.

The plateau values of the forward speed $v_{\rm f}$ are described with
good accuracy by the well-known simple linear decrease of the mean
speed $v_{\rm b}$ with mean density $\rho_{\rm b}$
\cite{fily2012prl,stenhammar2013,solon2015prl}, given by
\begin{align}
  \frac{v_{\rm b}}{s}=1-\frac{\rho_{\rm b}}{\rho_{\rm jam}},
  \label{EQvelocityLinearDecrease}
\end{align}
where $\rho_{\rm jam}$ is a constant that controls the slope of the
decrease of the mean swim speed in bulk, as well as the upper limit of
density (``jamming''). Here we take the convention that $\rho_{\rm b}$
indicates the number of particles per volume and per radians, hence
$2\pi\rho_{\rm b}$ is the number of particles per two-dimensional
volume.

In order to address the total force density balance
\eqref{EQforceDensityBalance}, we specify the internal force splitting
\eqref{EQforceDensitySplittingAdSup} and \eqref{EQFsupSplitting}
further by requiring that
\begin{align}
   -k_BT\nabla\rho + \Fv_{\rm ad} + \Fv_{\rm sup,3} &= 0,
  \label{EQforceDensityBalanceStructure}
\end{align}
which yields upon inserting into~\eqref{EQforceDensityBalance} the
relationship
\begin{align}
  \gamma \Jv &= \gamma s\ov\rho + \Fv_{\rm sup,0}
  + \Fv_{\rm sup,1} + \Fv_{\rm sup,2},
  \label{EQforceDensityBalanceMotion}
\end{align}
containing the motion (left hand side) and the contribution due to the
swimming (first term on the right hand side).  We will below identify
the superadiabatic force fields that determine via
\eqref{EQforceDensityBalanceMotion} the flow that occurs in the
system. Before doing so, in the following we first address the
structural force density balance
\eqref{EQforceDensityBalanceStructure}, {\mm which we will demonstrate
  to be a gradient relation when written in force field form.}

\subsection{Phase behaviour}

In order to address nonequilibrium phase
coexistence, we turn our description from force densities to force
fields. We hence divide \eqref{EQforceDensityBalanceStructure} by the
orientation- and position-resolved density distribution $\rho$, which
yields
\begin{align}
  - k_BT\nabla\ln\rho_0
  + \fv_{\rm ad}(\rv) 
  + \fv_{\rm sup,3}(\rv) 
  &= 0,
  \label{EQforceBalanceStructural}
\end{align}
where the adiabatic and superadiabatic force fields are defined by
$\fv_{\rm ad}=\Fv_{\rm ad}/\rho$ and $\fv_{\rm sup,3}=\Fv_{\rm
  sup,3}/\rho$, respectively. For simplicity, we have replaced in
\eqref{EQforceBalanceStructural} the ideal diffusion term
$-k_BT\nabla\ln\rho$ by the corresponding isotropic component
$-k_BT\nabla\ln\rho_0$. This is a good approximation, as we find that
the difference between isotropic and anisotropic gradients is a small
correction for relevant conditions, typically one or two orders
smaller in magnitude compared to all other contributions
\cite{footnote2}.

The adiabatic force field can be expressed as the negative gradient of
the local excess (over ideal gas) chemical potential, which within
classical density functional theory \cite{evans1979,evans2016} {\ms
  (see appendix \ref{SECdft} for a brief overview)} is given via
functional differentiation of the excess Helmholtz free energy
functional $F_{\rm exc}[\rho_0]$,
\begin{align}
  \fv_{\rm ad}(\rv) &=
  -\nabla \frac{\delta F_{\rm exc}[\rho_0]}{\delta \rho_0}.
  \label{EQfadDefinitionDFT}
\end{align}
Here the adiabatic state is an equilibrium system that possesses the
same density distribution as the real system. The real and the
adiabatic system share the same interparticle interaction
potential. The orientational degrees of freedom do not affect the
internal forces in equilibrium, as the particles are simple repulsive
disks. Hence, the free energy functional requires only the average
density profile $\rho_0$ as an input \cite{footnote4}, and $\fv_{\rm
  ad}$ is independent of orientations. (This situation is different in
a system of e.g.\ swimming rods or general anisotropic interparticle
interactions.) An equivalent alternative to the density functional
theory expression \eqref{EQfadDefinitionDFT} is $\fv_{\rm
  ad}=-\langle\nabla_i u(\rv^N)\rangle_{\rm eq}$, where $u(\rv^N)$ is
the interparticle interaction potential. Here the equilibrium average
is performed under the influence of a (hypothetical) ``adiabatic''
external potential $V_{\rm ad}(\rv)$, chosen such that the resulting
one-body density distribution $\rho$ is the correct one.  In practice,
this method requires to perform the average in e.g.\ Monte Carlo
simulations \cite{fortini14prl,delasheras2018customFlow}.

In order to describe the adiabatic force field, we use a simple local
density approximation, based on scaled-particle theory for
two-dimensional hard disks (although more accurate approximations
exist \cite{santos1995}),
\begin{align}
  \fv_{\rm ad} &= -\nabla \mu_{\rm ad},
  \label{EQfadAsGradient}
\end{align}
where the chemical potential $\mu_{\rm ad}$ for a bulk fluid of
density $\rho_{\rm b}$ is given by
\begin{align}
  \mu_{\rm ad}(\rho_{\rm b})&=
  k_BT\left[ -\ln(1-\eta') +
  \eta'\frac{3-2\eta'}{(1-\eta')^2}\right].
  \label{EQadiabaticMu}
\end{align}
Here we have introduced a rescaled packing fraction $\eta'=0.8\eta$ in
order to approximately take account of the repulsive sphere character
of the system; in our units $\eta=\rho_{\rm b}/\rho_{\rm jam}$.
Furthermore the bulk density $\bar\rho=N/V=2\pi\rho_{\rm b}$ indicates
the number of particles per unit volume.
The corresponding expression for the pressure can be obtained by
integrating the thermodynamical relation $\partial P_{\rm
  ad}/\partial\rho_{\rm b}=\rho_{\rm b}\partial \mu_{\rm ad}/\partial
\rho_{\rm b}$ in $\rho_{\rm b}$. This yields
\begin{align}
  P_{\rm ad}(\rho_{\rm b}) &= 
  k_BT\rho_{\rm b}\Big[\frac{1}{(1-\eta')^2}-1\Big].
  \label{EQadiabaticP}
\end{align}
{\ms The pressure generates the adiabatic force density, via a
  gradient operation, $ \rho_{\rm b}\fv_{\rm ad} = -\nabla P_{\rm ad}.
  $ The existence of the adiabatic force field and its corresponding
  integrals $\mu_{\rm ad}$ and $P_{\rm ad}$ is not that of an
  approximation. Rather this constitutes the part of the total
  nonequilibrium internal force density (and its corresponding
  position integrals) that is independent of velocity and hence
  dependent only on the density distribution (i.e.\ is a density
  functional). The ``genuine'' nonequilibrium contributions do also
  depend on the velocity field and are referred to as
  superadiabatic. We address the superadiabatic force fields in the
  following.

}

As $\fv_{\rm ad}$ is independent of orientation, $\fv_{\rm sup,3}$
also necessarily needs to be independent of $\ov$, in order to
satisfy~\eqref{EQforceBalanceStructural}. Furthermore, the adiabatic
force field is a gradient field, due to
\eqref{EQfadDefinitionDFT}. Hence in order for the force balance
\eqref{EQforceBalanceStructural} to hold, the superadiabatic force
field $\fv_{\rm sup,3}$ necessarily also needs to be of gradient form,
\begin{align}
  \fv_{\rm sup,3} &= -\nabla \nu_3,
  \label{EQfsup3AsGradient}
\end{align}
where $\nu_3(x)$ is the {\mm negative} spatial integral of the force field.

We first address the value of $\nu_3$ for constant density $\rho_{\rm
  b}$, i.e.\ far from the interface, where all gradients vanish.
Here we postulate an explicit form, which is quadratic in velocity,
given by
\begin{align}
  \nu_3(\rho_{\rm b}) &= e_1\frac{\gamma}{2D_{\rm rot}} v_{\rm b}^2
    \frac{\rho_{\rm b}}{\rho_{\rm jam}},
    \label{EQnu3}
\end{align}
where $e_1$ is a (dimensionless) constant that controls the strength
of the effect. Here {\mm our approximation \eqref{EQnu3} for} $\nu_3$
depends on density and velocity, but not directly on $s$, as is
consistent with the power functional framework
\cite{schmidt2013pft}. 
{\mm We find that using the form \eqref{EQnu3} we are able to satisfy
  the requirement that $\Fv_{\rm sup,3}$ is the remainder in the
  superadiabatic force splitting \eqref{EQFsupSplitting} in that no
  significant unexplained force contributions remain as compared to
  the simulation data. The magnitude of the observed numerical
  deviations is entirely consistent with the approximate nature of the
  expressions in our standalone theory.}

Using the explicit expression
\eqref{EQvelocityLinearDecrease} for $v_{\rm b}(\rho_{\rm b})$ allows
to obtain a corresponding pressure contribution $\Pi_3$ via
integration of $\partial \Pi_3/\partial \rho_{\rm b} = \rho_{\rm
  b}\partial \nu_3/\partial \rho_{\rm b}$.  The result is
\begin{align}
  \Pi_3(\rho_{\rm b}) &=
  \frac{\gamma e_1}{4D_{\rm rot}\rho_{\rm jam}}
  v_{\rm b}^2\rho_{\rm b}^2
  \left[1+
  \frac{\rho_{\rm b}(3\rho_{\rm b}-4\rho_{\rm jam})}
       {6(\rho_{\rm jam}-\rho_{\rm b})^2}
       \right].
  \label{EQPi3}
\end{align}

{\mz In order to obtain the quantities that determine nonequilibrium
  phase coexistence, we return to the force balance relation
  \eqref{EQforceBalanceStructural}, which we rewrite using the
  gradient expressions \eqref{EQfadAsGradient} and
  \eqref{EQfsup3AsGradient} as
\begin{align}
  -\nabla(k_BT\ln\rho_0 
  +\mu_{\rm ad} 
  +\nu_3) &= 0.
\end{align}
As the total gradient vanishes, the expression in brackets needs to be
equal to a constant, $\mu=\rm const$, which plays the role of the
total chemical potential. Similarly, the force density balance
relation \eqref{EQforceDensityBalanceStructure} can be rewritten as
\begin{align}
  -\nabla(k_BT\rho_0 + P_{\rm ad} + \Pi_3) &= 0,
  \label{EQpressureGradient}
\end{align}
which again implies that the expression in brackets is constant, where
the constant, $P=\rm const$, plays the role of the total pressure.}
{\mm The three individual terms inside of the gradient on the left
  hand side of \eqref{EQpressureGradient} depend in general on
  position $x$. As before, $\rho_0(x)$ is the isotropic Fourier
  component of the density profile.}

We can now define bulk values of the total chemical potential $\mu$
and the total pressure $P$ by summing up the individual contributions,
\begin{align}
  \mu(\rho_{\rm b}) &= k_BT\ln\rho_{\rm b} + \mu_{\rm ad} + \nu_3,
  \label{EQtotalChemicalPotentialDefinition3ad}\\
  P(\rho_{\rm b}) &= k_BT\rho_{\rm b} + P_{\rm ad} + \Pi_3.
  \label{EQtotalPressureDefinition3ad}
\end{align}
We show below that although further contributions to the chemical
potential exist, the sum of these additional contributions
vanishes. Hence \eqref{EQtotalChemicalPotentialDefinition3ad} indeed
defines the total chemical potential. The same holds true for the
total pressure, where we demonstrate below that although there is a
swim pressure contribution, this is identically cancelled by a
corresponding superadiabatic (i.e.\ intrinsic nonequilibrium) pressure
that the system develops. Hence \eqref{EQtotalPressureDefinition3ad}
represents the total pressure in nonequilibrium bulk steady states.

Phase coexistence implies that the densities in the coexisting gas and
liquid phases, $\rho_{\rm g}$ and $\rho_{\rm l}$, respectively,
satisfy
\begin{align}
  \mu(\rho_{\rm g}) &= \mu(\rho_{\rm l}),\label{EQcoexistenceMu}\\
  P(\rho_{\rm g}) &= P(\rho_{\rm l}).\label{EQcoexistencePressure}
\end{align}
Equations \eqref{EQcoexistenceMu} and \eqref{EQcoexistencePressure},
together with \eqref{EQtotalChemicalPotentialDefinition3ad} and
\eqref{EQtotalPressureDefinition3ad}, 
\eqref{EQadiabaticMu} and \eqref{EQadiabaticP}, and
\eqref{EQnu3} and \eqref{EQPi3}, form a closed set of equations for the
determination of the binodal densities $\rho_{\rm g}$ and $\rho_{\rm
  l}$, which we solve numerically.  Figure~\ref{figPhaseDiagram}
presents the theoretical results for the phase diagram, and comparison
to simulation data from the literature. Here we have chosen
$e_1=0.0865$ and $\rho_{\rm jam}2\pi\sigma^2 =1.146$. The phase
diagram possesses a lower, in Peclet number ${\rm Pe}\equiv 3s/(\sigma
D_{\rm rot})=\gamma s \sigma/(k_BT)$, critical point (which is
characterized by mean-field exponents within the present approach; see
Ref.~\cite{binder2018} for a simulation study of the critical
scaling.) The binodal agrees very well with the simulation data of
Ref.~\cite{paliwal2018njp}.  We obtain the spinodal via the condition
$\partial\mu(\rho_{\rm b})/\partial \rho_{\rm b} = 0.$ The critical
point is then obtained by the additional condition
$\partial^2\mu/\partial \rho_{\rm b}^2 = 0$. This necessitates finding
the appropriate root of a fourth-order polynomial in the value of the
critical density \cite{footnote3}. We perform this task numerically.

We show in Fig.~\ref{figPhaseDiagram} the theoretical result for the
spinodal density, together with the simulation data for the gas side
of the spinodal by Stenhammar et al.~\cite{stenhammar2013}. Clearly,
the agreement between the theoretical results and the simulation data
is very satisfactory. The theory in particular captures correctly the
fact that the gas side of both the spinodal and the binodal remain at
relatively large density upon increasing $\rm Pe$. This is in striking
contrast to typical equilibrium gas-liquid coexistence, where the gas
becomes rapidly very dilute upon increasing distance (in temperature)
from the critical point.

Having established the bulk phase diagram, in the following we develop
a microscopic theory for the interfacial structure between coexisting
active gas and active liquid states. This allows us to demonstrate (i)
that the sum of all further contributions to the state functions
$\mu(\rho_{\rm b})$ and $P(\rho_{\rm b})$ indeed vanishes, and hence
that \eqref{EQtotalChemicalPotentialDefinition3ad} and
\eqref{EQtotalPressureDefinition3ad} are complete, and (ii) that bulk
coexistence is unaffected by interfacial contributions.

\subsection{Fourier decomposition}

We restrict ourselves to steady states of two-dimensional systems,
which are spatially inhomogeneous only in the $x$-direction;
orientation is measured by the angle $\varphi$ against the $x$-axis,
i.e.\ $\ov=(\cos\varphi,\sin\varphi)$. We Fourier decompose the
kinematic fields $\rho$ and $\Jv$ according to
\begin{align}
  \rho(x,\varphi) &= \sum_{n=0}^\infty \rho_n(x) \cos(n\varphi),
  \label{EQFourierDensity}\\
  J_x(x,\varphi) &= \sum_{n=1}^\infty J_n^{x}(x) \cos(n\varphi),
  \label{EQFourierCurrentx}\\
  J_y(x,\varphi) &= \sum_{n=1}^\infty J_n^{y}(x) \sin(n\varphi),
  \label{EQFourierCurrenty}
\end{align}
where the Cartesian components of the one-body current are
$(J_x,J_y)\equiv\Jv$ and $\rho_n$, $J_n^x$, $J_n^y$ are Fourier
coefficients which depend on position $x$. Terms that vanish due to
symmetry in $\varphi$ have been omitted: the system is invariant under
reflection {\mb with respect to the $x$-axis}, i.e.\ under the joint
coordinate transformation $y\to-y$ and $\varphi\to-\varphi$. Hence the
density \eqref{EQFourierDensity} and the $x$-component of the current
\eqref{EQFourierCurrentx} need to be even in $\varphi$; the
$y$-component of the current \eqref{EQFourierCurrenty} flips its
direction under the reflection, and hence it is odd in~$\varphi$.
Furthermore, as is common, we restrict ourselves to cases where the
isotropic current component vanishes, $J_0^x=J_0^y=0$. Using the
low-order Fourier coefficients $n=0,1$, we can express frequently used
standard observables: the orientation-integrated density distribution
is simply $2\pi\rho_0$ and the polarization profile with respect to
the $x$-axis is $\pi\rho_1$.  Finally, we use the convention that
$x=0$ indicates the position of the Gibbs dividing surface {\mm
  \cite{Hansen06}}.

In the present case the rotational derivative $\nabla^\omega$ is
simply $\partial/\partial\varphi$ and the general continuity equation
\eqref{EQcontinuity} reduces to
\begin{align}
  \frac{\partial J_x}{\partial x} &=
  D_{\rm rot} \frac{\partial^2 \rho}{\partial\varphi^2}.
\end{align}
Upon inserting the Fourier ansatz \eqref{EQFourierDensity} and
\eqref{EQFourierCurrentx}, this can be cast into a relationship for
the Fourier coefficients of the density of order $n\geq 1$,
\begin{align}
  \rho_n &= -\frac{1}{n^2D_{\rm rot}} \frac{d J_n^x}{dx},
  \label{EQdensityCoefficients}
\end{align}
which we will use below in order to derive a set of (coupled)
differential equations for the Fourier coefficients.

\subsection{Drag forces}

In order to address the dynamical force balance
\eqref{EQforceDensityBalanceMotion}, we specify the superadiabatic
contributions in \eqref{EQFsupSplitting} further, by requiring that
the drag force density $\Fv_{\rm sup,0}$ acts against the flow
direction, and is given by an orientational average, defined by the
correlator
\begin{align}
  \Fv_{\rm sup,0} &= \frac{\Jv}{2\pi J_{\rm f}} 
  \int d\ov' \ov' \cdot \Fv_{\rm int}(\rv,\ov').
  \label{EQFsup0PrimeAsCorrelator}
\end{align}
where the local orientation-averaged forward current profile $J_{\rm
  f}(x)$ is defined via \eqref{EQJfAsAverage} {\ms and $\ov'$ is a new
  angular integration variable. In the present two-dimensional system
  the integration over orientation space is simply $\int d\ov\equiv
  \int_{-\pi}^\pi d\varphi$. } The force density field $\Fv_{\rm
  sup,0}(\rv,\ov)$ depends on orientation $\ov$ via the dependence of
$\Jv$ on $\ov$ on the right hand side.

In order to develop the theory, we assume the drag force density
\eqref{EQFsup0PrimeAsCorrelator} to have the form
\begin{align}
  \Fv_{\rm sup,0} &= -
  \frac{\gamma\rho_0}{\rho_{\rm jam}-\rho_0}
       [1+\xi(\nabla\rho_0)^2]\Jv,
  \label{EQFsup0PrimeAsFunctional}
\end{align}
where $\xi>0$ is a constant (with units of ${\rm length}^2/{\rm
  density}^2$) that determines the strength of the square gradient
correction. For the case of constant density, $\nabla\rho_0=0$, the
expression \eqref{EQFsup0PrimeAsFunctional} reduces to the previously
formulated bulk fluid drag force \cite{krinninger2016prl}, which
reproduces the well-known
\cite{fily2012prl,stenhammar2013,solon2015prl} linear decrease
\eqref{EQvelocityLinearDecrease} of the mean speed with increasing
average density in bulk. Equation \eqref{EQFsup0PrimeAsFunctional}
constitutes a kinematic functional (i.e.\ the dependence is on $\rho$
and $\Jv)$, as required \cite{schmidt2013pft,krinninger2016prl}.

In order to describe the orientation-averaged density profile across
the interface, we use the classic form
\begin{align}
  \rho_0(x) &= \frac{\rho_{\rm l}+\rho_{\rm g}}{2} 
  + \frac{\rho_{\rm l}-\rho_{\rm g}}{2}\tanh(x/\lambda),
  \label{EQrho0Tanh}
\end{align}
where the lengthscale $\lambda$ determines the width of the interface.
The corresponding mean densities, with units of particle number per
system volume, are $2\pi\rho_{\rm g}$ and $2\pi\rho_{\rm l}$.
Equation~\eqref{EQrho0Tanh} is widely used in the description of the
present problem \cite{paliwal2018njp,paliwal2017jcp} and it is
considered to be an excellent approximation to simulation results.

We next identify the non-spherical drag correction, defined by
\begin{align}
  \Fv_{\rm sup,1} &= \frac{\ov^*}{2\pi}
  \int d\ov' 
  (\Fv_{\rm int}'-\Fv_{\rm sup,0}' )\cdot\ov'^*,
  \label{EQFsup1PrimeAsAverage}
\end{align}
where $\ov^*=(\cos\varphi,-\sin\varphi)$ is the orientation $\ov$
reflected at the $x$-axis. (Note that when viewing the set of
$xy$-coordinates as the complex plane, then $\ov^\ast$ is the complex
conjugate to $\ov$.) {\ms Furthermore, the primed force density fields
  inside of the orientation integral are evaluated at direction
  $\ov'$.}  In the theory, we postulate the nonspherical drag to have
the form \cite{footnote5}:
\begin{align}
  \Fv_{\rm sup,1} &= 
  -\frac{\gamma}{4} \rho_1 v_{\rm f}
  \frac{1+\xi(\nabla\rho_0)^2\rho_0/\rho_{\rm jam}}
       {1-\rho_0/\rho_{\rm jam}}\ov^*,
       \label{EQFsup1PrimeAsFunctional}
\end{align}
{\mm which is linear in the forward speed $v_{\rm f}$, as is
  appropriate for a drag term, and contains a square density gradient
  contribution to its amplitude. The direction of the nonspherical
  drag is against the $\ov^\ast$ direction.}

\begin{figure}
  \includegraphics[width=0.46\textwidth]{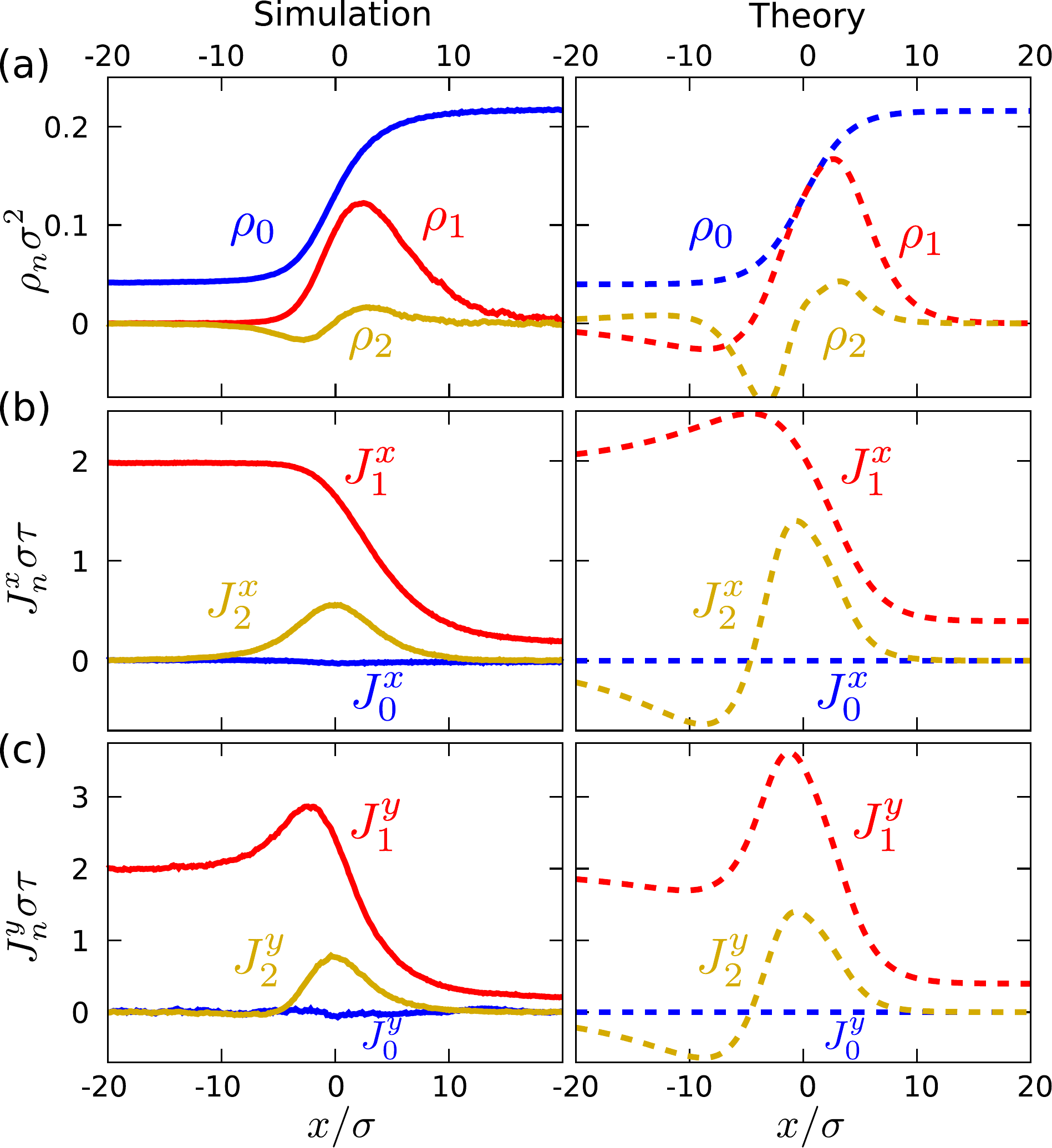}
  \caption{Representative Fourier coefficients obtained from BD
    simulation (left panels) and theory (right panels). Shown are
    Fourier components of order $n=0,1,2$ (as indicated) as a function
    of $x/\sigma$ for (a) the density $\rho_n\sigma^2$, (b) the
    $x$-component of the current, $J_n^x\sigma\tau$, and (c) the
    $y$-component of the current,~$J_n^y\sigma\tau$, where the
    ``molecular'' timescale is $\tau=\gamma \sigma^2/\epsilon$. The
    simulation parameters are identical to those of Fig.~\ref{fig1}.}
  \label{figFourier}
\end{figure}

\begin{figure*}
  \includegraphics[width=0.95\textwidth]{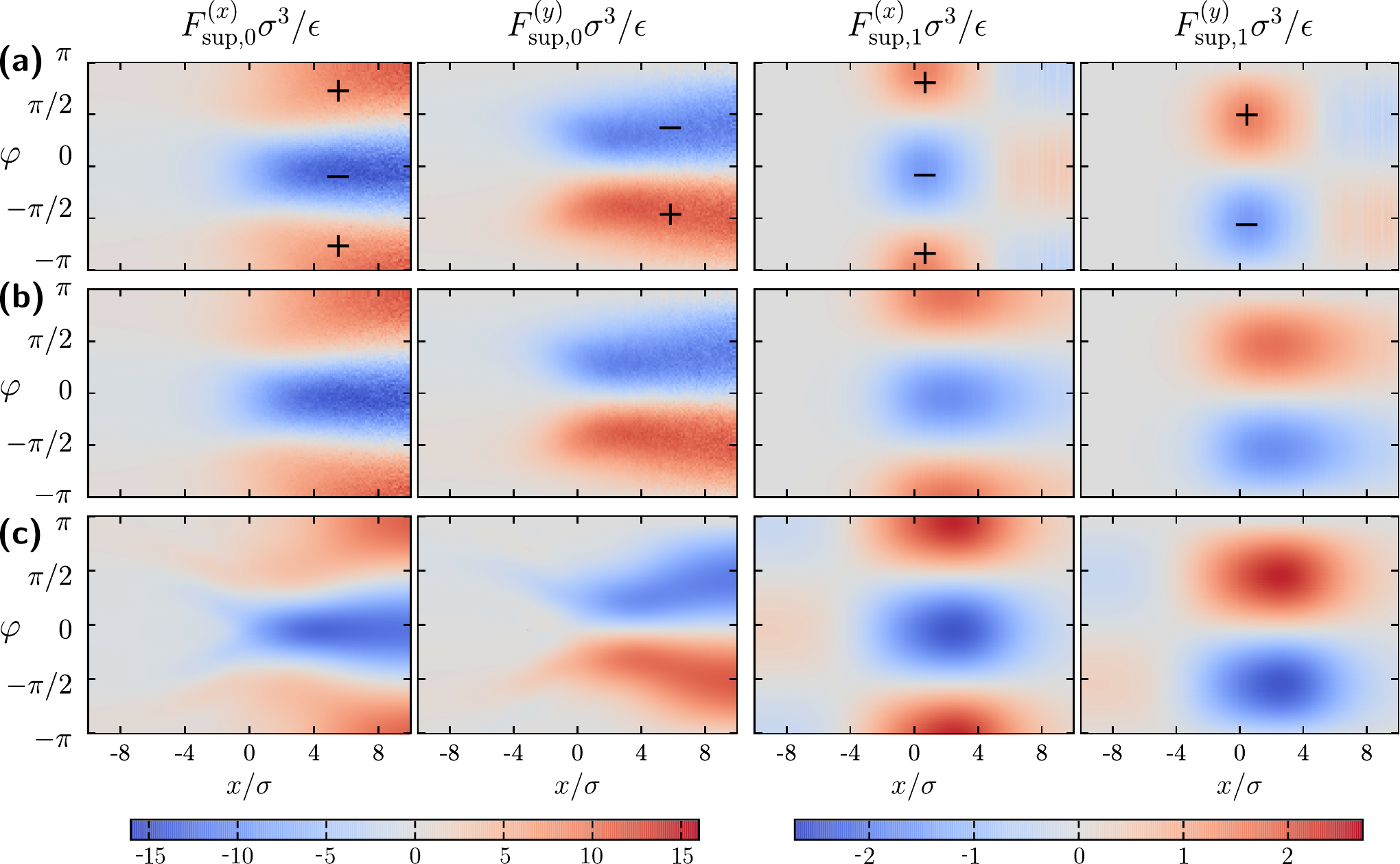}
  \caption{Spherical drag force density $\Fv_{\rm sup,0}$ and
    nonspherical drag force density $\Fv_{\rm sup,1}$, obtained from
    (a) simulations via correlators, (b) via kinematic functionals
    using simulation data as input, and (c) stand-alone theory. Shown
    are the $x-$ and $y-$components of the respective force density
    field: $F_{\rm sup,0}^{(x)}$ (first column) $F_{\rm sup,0}^{(y)}$
    (second column), $F_{\rm sup,1}^{(x)}$ (third column), and $F_{\rm
      sup,1}^{(y)}$ (fourth column) in units of $\epsilon/\sigma^3$
    and as a function of distance $x/\sigma$ across the interface and
    angle $\varphi$ with respect to the interface normal (pointing
    towards the liquid). The plus and minus signs indicate the sign of
    the force density fields.}
\label{figForceFields}
\end{figure*}

\begin{figure*}
  \includegraphics[width=0.9\textwidth]{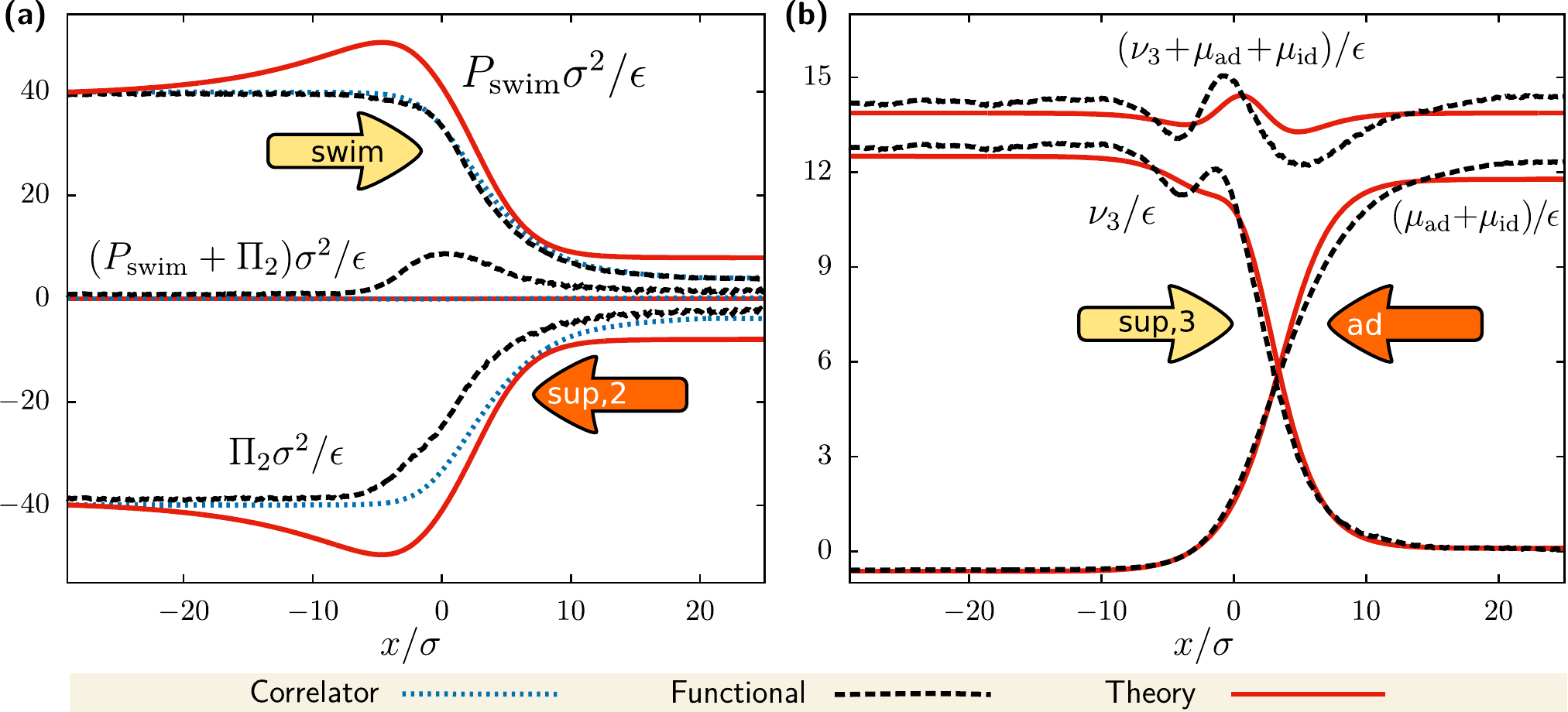}
  \caption{(a) Superadiabatic spherical pressure profile $\Pi_2(x)$,
    swim pressure profile $P_{\rm swim}(x)$, and the sum
    $\Pi_2(x)+P_{\rm swim}$, in units of $\epsilon/\sigma^2$ as a
    function of distance $x/\sigma$ across the interface. Results are
    obained from the correlator expressions using simulation data as
    input (blue dotted lines), from the kinematic functionals with
    simulation data input (black dashed lines) and from the stand
    alone theory (red solid lines). (b) Superadiabatic spherical
    chemical potential profile $\nu_3(x)$, the sum of adiabatic excess
    and ideal chemical potential profile $\mu_{\rm id}(x)+\mu_{\rm
      ad}(x)$, and the total chemical potential, i.e.\ the sum
    $\nu_3+\mu_{\rm ad}+\mu_{\rm id}$ in units of $\epsilon$ and as a
    function of $x/\sigma$. Here the ideal chemical potential is
    $\mu_{\rm id}=k_BT \ln(\eta')$ with the local rescaled packing
    fraction $\eta'=0.8\rho_{\rm b}/\rho_{\rm jam}$. For clarity the
    results for the sum have been shifted upwards by two units. The
    superadiabatic results are obtained using the kinematic functional
    with either the simulated (dashed black lines) or stand alone
    (solid red lines) Fourier coefficients.}
\label{figPotentials}
\end{figure*}

We next insert \eqref{EQFsup0PrimeAsFunctional} and
\eqref{EQFsup1PrimeAsFunctional} into
\eqref{EQforceDensityBalanceMotion}, and express the kinematic fields
using their Fourier forms \eqref{EQFourierDensity},
\eqref{EQFourierCurrentx} and \eqref{EQFourierCurrenty}. {\mm (The
  contribution $\Fv_{\rm sup,2}$ will be considered below in
  Sec.~\ref{SECsphericalPressure}.)} The factor
$\ov=(\cos\varphi,\sin\varphi)$ that occurs in the swim force density
couples the different modes. Using trigonometric identities allows to
rearrange all expressions into a single Fourier series. Satisfying the
force density balance \eqref{EQforceDensityBalanceMotion} is then
equivalent to requiring that the prefactor of each mode $n$
vanishes. Together with Eqs.~\eqref{EQJfAsAverage} and
\eqref{EQvfAsAverage} this leads to the coupled set of algebraic
relations
\begin{align}
  J_1^x = v_{\rm f}\Big(\rho_0-\frac{\rho_1}{4}+
  \frac{\rho_2}{2}\Big),\quad
  J_{n>1}^x &= \frac{v_{\rm f}}{2}(\rho_{n-1}+\rho_{n+1}),
  \label{EQJnx}\\
  \quad J_1^y = v_{\rm f}\Big(\rho_0+\frac{\rho_1}{4}
  -\frac{\rho_2}{2}\Big),\quad
  J_{n>1}^y &= \frac{v_{\rm f}}{2}(\rho_{n-1}-\rho_{n+1}).
  \label{EQJny}
\end{align}
Here the prefactor $v_{\rm f}$ is the forward speed profile, given as
\begin{align}
  v_{\rm f} &= s
  \frac{1-\rho_0/\rho_{\rm jam}}
       {1+\xi(\nabla\rho_0)^2\rho_0/\rho_{\rm jam}}.
  \label{EQvfAsFunctional}
\end{align}
As a special case, in the homogeneous isotropic bulk, the density
gradient vanishes, and \eqref{EQvfAsFunctional} reduces to
\eqref{EQvelocityLinearDecrease}.

We are now in a position to compare {\mm results} for the structure
from simulation and from theory quantitatively. Figure~\ref{fig1}(d)
shows {\mm results} for the forward speed obtained from simulations
via \eqref{EQvfAsAverage} against the representation
\eqref{EQvfAsFunctional} (see inset).
We have set the parameters $\xi=700
\sigma^6\approx (3\sigma)^6$, cf.~\eqref{EQvfAsFunctional}, and
$\rho_{\rm jam}2\pi\sigma^2 =1.4$ in order to best match theoretical
and simulation data; we keep these values for all further
comparisons. 
{\mm (Here we have readjusted the value of $\rho_{\rm jam}$, because the
  control parameter in our simulations are different from those of
  Ref.~\cite{paliwal2018njp}.)}
Here the theoretical result is taken at nonequilibrium coexistence, and
the average density profile $\rho_0$ is in steady state.  The theory
correctly describes the smooth crossover from the fast motion in the
gas to the slow motion in the liquid.

Replacing $J_n^x$ via \eqref{EQJnx} in the relationship of density and
current coefficients \eqref{EQdensityCoefficients}, yields a closed
set of coupled first-order ordinary differential equations for the
coefficients $\rho_n$, given by
\begin{align}
  -D_{\rm rot} \rho_1 &= \frac{d}{dx} v_{\rm f}
  \left(\rho_0-\frac{\rho_1}{4}+\frac{\rho_2}{2}\right),
  \label{EQrecursionStart}\\
  - D_{\rm rot} \rho_{n} &= \frac{d}{dx} \frac{v_{\rm f}}{2n^2}
  \left(\rho_{n-1} + \rho_{n+1}\right), \qquad n>1.
  \label{EQrecursionStep}
\end{align}
Once the $\rho_n$ are known, one can (trivially) determine the $J_n^x$
and $J_n^y$ via \eqref{EQJnx} and \eqref{EQJny}. As an aside, note
that the sum rule $2 v_{\rm f}\rho_0=J_1^x+J_1^y$, which can be
derived from inserting \eqref{EQFourierCurrentx} and
\eqref{EQFourierCurrenty} into \eqref{EQvfAsAverage}, is satisfied by
\eqref{EQJnx} and \eqref{EQJny}.  Note also that the coupling of the
orientational and the translational motion occurs now (only) via the
shifted indices $n\pm 1$ in \eqref{EQJnx} and \eqref{EQJny}. We are
now at the stage that the force density balance
\eqref{EQforceDensityBalanceMotion} is satisfied at all positions
across the interface and for all orientational modes that are present
in the system (i.e.\ for $n\geq 1$).

In order to construct an approximative explicit solution, we neglect
both $\rho_2$ in \eqref{EQrecursionStart} and $\rho_{n+1}$ in
\eqref{EQrecursionStep}. This then allows to obtain all $\rho_n(x)$
numerically by simple iteration, starting with $\rho_0(x)$ given by
\eqref{EQrho0Tanh}.  We show a comparison of the agreement of the
Fourier coefficients obtained from this theory and from simulations in
Fig.~\ref{figFourier}(a) for $\rho_n$, in Fig.~\ref{figFourier}(b) for
$J_n^x$, and in Fig.~\ref{figFourier}(c) for $J_n^y$. The theory
captures all qualitative features of the simulation data with a slight
tendency for over-structuring. We attribute the small overshoot
effects to the truncation of the full recursion relation. The theory
in particular describes the polarization of the interface (peak in
$\rho_1$), as well as the oscillating structuring of the ``nematic''
order as measured by $\rho_2$. The $x$-component of the current shows
a decay of the primary component, $J_1^{x}$, when traversing from the
gas to the liquid phase. Again the next higher Fourier component,
$J_2^x$ is peaked at the interface (as is $J_2^y$). The $y-$component
of the current, $J_1^y$ measures flow parallel to the interface. This
component has the same bulk plateau values as the corresponding
$x-$component, but shows a pronounced peak at the interface. In
particular this effect is well described by the theory. Furthermore,
within the approximative solution the strict equality $J_n^x=J_n^y$
holds for $n\geq 2$. The simulation data (compare left panels of
Fig.~\ref{figFourier} (b) and (c)) indicates that this is indeed a
reasonable approximation for $n=2$. {\mb See appendix
  \ref{SECboxGeometry} for a description of the influence of the box
  geometry on the simulation results. Appendix
  \ref{SECsquareGradientParameter} describes the effects of changing
  the value of the square gradient parameter~$\xi$.}

Figure~\ref{figForceFields}(a) displays results for the spherical drag
force density profile $\Fv_{\rm sup,0}$ as a function of distance $x$
across the interface and angle $\varphi$ with respect to the interface
normal (recall $\varphi=0$ corresponds to the direction towards the
liquid). We show simulation results from using the correlator
\eqref{EQFsup0PrimeAsCorrelator} applied to the ``raw'' simulation
data for $\Fv_{\rm int}$. In Fig.~\ref{figForceFields}(b) we show
results obtained from the kinematic expression
\eqref{EQFsup0PrimeAsFunctional} and using the simulation results for
$\rho_0$ and $\Jv$ as input. The agreement with the results from the
correlator expressions shown in Fig.~\ref{figForceFields}(a) is
impressive and validates the form \eqref{EQFsup0PrimeAsFunctional} of
the spherical drag force. We also compare against results from the
stand-alone theory, where we use the kinematic expression
\eqref{EQFsup0PrimeAsFunctional}, the ansatz for the density profile
\eqref{EQrho0Tanh}, the result of the truncated hierarchy of Fourier
coefficients, and the kinematic expression \eqref{EQvfAsFunctional}
for $v_{\rm f}$. These theoretical results are shown at bulk
coexistence, which fixes the values for the coexisting densities
$\rho_{\rm g},\rho_{\rm l}$ and for the interfacial width $\lambda$.
Although some artifacts occur, the stand-alone theory describes the
isotropic drag force field quantitatively correctly,
cf.~Fig.~\ref{figForceFields}(c).

Results for the nonspherical drag force density field $\Fv_{\rm
  sup,1}$ are shown in the third and in the fourth column of
Fig.~\ref{figForceFields}. We use the same three types of approaches
as above.  In Fig.~\ref{figForceFields}(a) results are shown from the
correlator \eqref{EQFsup1PrimeAsAverage} applied to the simulation
data.  Fig.~\ref{figForceFields}(b) presents the results from the
kinematic expression \eqref{EQFsup1PrimeAsFunctional} applied to the
simulation data.  Fig.~\ref{figForceFields}(c) presents results from
the stand-alone theory using the kinematic expression
\eqref{EQFsup1PrimeAsFunctional} with approximated Fourier
coefficients. The agreement between all three approaches is again
excellent, with small artifacts displayed by the stand-alone theory.
The spherical drag force (first and second column) indeed opposes the
motion. Its magnitude is small in the gas and large in the liquid, and
it crosses over continuously between these limits. The nonspherical
contribution (third and fourth colums) is qualitatively similar, but
acts only in the interfacial region. Its magnitude is smaller by more
than a factor of 5 than that of $\Fv_{\rm sup,0}$.

\subsection{Spherical superadiabatic pressure}
\label{SECsphericalPressure}
We specify the
superadiabatic force density field via
\begin{align}
  \Fv_{\rm sup,2} &= 
  \frac{\ev_x}{2\pi}\int d\ov \Fv_{\rm int}\cdot \ev_x.
  \label{EQFsup2AsCorrelator}
\end{align}
Per construction $\Fv_{\rm sup,2}$ is independent of
orientation. Hence considering the isotropic mode, $n=0$, of the force
density balance \eqref{EQforceDensityBalanceMotion} allows one to
identify $\Fv_{\rm sup,2}$ as a gradient expression,
\begin{align}
  \Fv_{\rm sup,2} &= -\nabla \Pi_2,
  \label{EQFsup2AsGradient}
\end{align}
where $\Pi_2$ is a superadiabatic spherical one-body pressure
contribution, which originates from the (repulsive) interparticle
interactions. {\mm From observing the gradient structure in
  \eqref{EQFsup2AsGradient}} we find its form to be
\begin{align}
  \Pi_2 &=
  -\frac{\gamma v_{\rm f}v_{\rm loc}}{2D_{\rm rot}}
  \frac{\rho_0 -\rho_1/4+\rho_2/2}{1-\rho_0/\rho_{\rm jam}},
  \label{EQsphericalPressure}
\end{align}
where we have defined
\begin{align}
  v_{\rm loc} &= v_{\rm f}\Big[1+\xi\frac{\rho_0}{\rho_{\rm jam}}
    (\nabla\rho_0)^2\Big],
  \label{EQvlocDefinition}
\end{align}
with the forward speed $v_{\rm f}$ being given by
\eqref{EQvfAsAverage}. (Here we have neglected the ideal diffusion
term \cite{footnote2}.) The spherical pressure depends on the density
and velocity fields, and is hence a kinematic functional, as expected
from \eqref{EQFsup2AsCorrelator} and \eqref{EQFsup2AsGradient}. As we
demonstrate below, the spherical pressure is negative and its
magnitude is high in the liquid and low in the gas.

Furthermore, there occurs a swim pressure contribution $P_{\rm swim}$,
which is due to the polarization of the interface. The corresponding
force density $\Fv_{\rm swim}$ and the swim pressure $P_{\rm swim}$
are defined, analogously to \eqref{EQFsup2AsCorrelator} and
\eqref{EQFsup2AsGradient} as 
\begin{align}
  \Fv_{\rm swim} &= \frac{\ev_x}{2\pi}
  \int d\ov \rho \gamma s \ov \cdot {\ms \ev_x},
  \label{EQFswimAsCorrelator}\\
  \Fv_{\rm swim} &= -\nabla P_{\rm swim}.
  \label{EQFswimAsGradient}
\end{align}
where the integrand in \eqref{EQFswimAsCorrelator} is the swim force
density, cf.~\eqref{EQforceDensityBalance}.
From inserting the Fourier series \eqref{EQFourierDensity} for $\rho$
into \eqref{EQFswimAsCorrelator} and using \eqref{EQvfAsFunctional}
and \eqref{EQrecursionStart} we obtain the swim pressure as
\begin{align}
  P_{\rm swim} &= \frac{\gamma s^2}{2D_{\rm rot}}
  \Big(\rho_0 - \frac{\rho_1}{4} + \frac{\rho_2}{2}\Big)
  \frac{1-\rho_0/\rho_{\rm jam}}{1+\xi(\nabla\rho_0)^2\rho_0/\rho_{\rm jam}}.
  \label{EQPswimAsExternal}
\end{align}
As expected from \eqref{EQFswimAsCorrelator} the swim pressure
\eqref{EQPswimAsExternal} depends on the free swim speed $s$ and on
the density distribution (via its angular Fourier coefficient
profiles).  A priori there is no dependence of $P_{\rm swim }$ on the
velocity field, and hence $P_{\rm swim}$ constitutes a density
functional, which parametrically depends on $s$.  The absence of a
dependence on the velocity field is again consistent with power
functional theory, as only intrinsic (superadiabatic) force density
fields, and hence their integrals, possess this dependence.

We can algebraically simplify the expression \eqref{EQPswimAsExternal}
for $P_{\rm swim}$ by using \eqref{EQvfAsFunctional} in order to
replace one factor of $s$. This yields the more compact form
\begin{align}
  P_{\rm swim} &= \frac{\gamma s v_{\rm f}}{2D_{\rm rot}} 
  \Big(\rho_0-\frac{\rho_1}{4}+\frac{\rho_2}{2}\Big).
  \label{EQPswimOnevf}
\end{align}
For bulk fluids the Fourier components $\rho_1=\rho_2=0$, and
\eqref{EQPswimOnevf} reduces to the previously obtained
\cite{solon2015prl,yang2014sm,takatori2014prl,solon2018njp} result
$P_{\rm swim}=\gamma s v_{\rm f}(\rho_0)\rho_0/(2D_{\rm rot})$.  This
is an important result and demonstrates that our strategy of working
on the level of force balance relationships does indeed describe the
correct physics.

By inserting the expression for the local speed
\eqref{EQvlocDefinition} into \eqref{EQsphericalPressure} and using
\eqref{EQvfAsFunctional} we can obtain an expression for $\Pi_2$,
which is up to a minus sign identical to the right hand side of
\eqref{EQPswimOnevf}. Hence we find that the swim pressure and the
spherical superadiabatic pressure cancel each other,
\begin{align}
  P_{\rm swim} + \Pi_2 &= 0.
  \label{EQpressureBalance2}
\end{align}
As the sum of the additional pressure contributions vanishes, there is
no effect on the total pressure \eqref{EQtotalPressureDefinition3ad},
and hence no influence on phase coexistence.

Eliminating the remaining dependence of $v_{\rm f}$ in favour of
dependence on $\rho_{\rm b}$ via the same procedure yields for bulk
fluid states
\begin{align}
  \Pi_{\rm 2}(\rho_{\rm b}) &= -\frac{\gamma s^2}{2D_{\rm rot}}
  \rho_{\rm b}\left(1-\frac{\rho_{\rm b}}{\rho_{\rm jam}}\right),\\
  \nu_2(\rho_{\rm b}) &= \frac{\gamma s^2}{2D_{\rm rot}}
  \left( \frac{2\rho_{\rm b}}{\rho_{\rm jam}}-\ln\rho_{\rm b}
  \right),
\end{align}
where $\nu_2$ acts as a bulk chemical potential contribution
corresponding to $\Pi_2$.  It is straightforward to obtain the
corresponding chemical potential for the swim contribution, and
\begin{align}
  \mu_{\rm swim} + \nu_2 &= 0.
  \label{EQmunu2vanishes}
\end{align}
Again, the total chemical potential
\eqref{EQtotalChemicalPotentialDefinition3ad} and hence the phase
behaviour is unaffacted by adding both additional chemical potential
contributions, as their sum \eqref{EQmunu2vanishes} vanishes.  Clearly
both $\Pi_2$ and $\nu_2$, as well as $P_{\rm swim}$ and $\mu_{\rm
  swim}$, constitute nonequilibrium bulk state functions for the
system.

Figure~\ref{figPotentials}(a) displays results for the intrinsic
spherical pressure profile and for the swim pressure profile across
the interface. Here we apply the correlator
\eqref{EQFsup2AsCorrelator} for $\Fv_{\rm sup,2}$ to the simulation
data for the density distribution $\rho$. Integrating
\eqref{EQFsup2AsGradient} in position then yields benchmark results
for the pressure profile $\Pi_2(x)$. Furthermore, we apply the
kinematic expression \eqref{EQsphericalPressure} to the simulation
results for the Fourier coefficients $\rho_0, \rho_1$, and $\rho_2$.
Thirdly, using the approximate form of the $\rho_n$ provides
stand-alone theoretical results for $\Pi_2(x)$.

We apply the same concept to the swim pressure. Here the benchmark
results are obtained by applying the correlator
\eqref{EQFswimAsCorrelator} for $\Fv_{\rm swim}$ to the simulation
data and integrating \eqref{EQFswimAsGradient} in position in order to
obtain $P_{\rm swim}$. Furthermore we take the expression
\eqref{EQPswimOnevf} with either simulation data as input or with
stand-alone Fourier coefficents as input.

In the stand-alone theory the total pressure profile vanishes
identically. Using the correlator expressions on the simulation data
confirms this result within the numerical precision. The functional
expressions generate a small (positive) artifact at the
interface. This is due to the (small) disagreement of $v_{\rm f}$, as
given by the analytical expression \eqref{EQvfAsFunctional}, with the
simulation result, cf.~Fig.~\ref{fig1}(d).  {\ms The defect of the
  theory giving non-vanishing values of $P_{\rm swim}+\Pi_2$ in the
  interfacial region can be traced back to the approximate nature of
  the relation \eqref{EQvfAsFunctional} of the forward speed $v_{\rm
    f}$ with the density profile. The insufficient cancellation can be
  (formally) avoided by replacing $s$ in \eqref{EQPswimAsExternal} by
  \eqref{EQvfAsFunctional}, such that the error cancels, and the sum
  of the pressures \eqref{EQsphericalPressure} and
  \eqref{EQPswimAsExternal} hence vanishes.}

Note that the superadiabatic contributions
\eqref{EQFsup0PrimeAsFunctional}, \eqref{EQFsup1PrimeAsFunctional},
and \eqref{EQFsup2AsGradient}, render the force density balance
\eqref{EQforceDensityBalanceMotion} to be satisfied at all positions
$x$, irrespective of the values of $\rho_g, \rho_l$ and $\lambda$.

\subsection{Quiet life force profile}
We have by now established both that the motional force density
balance \eqref{EQforceDensityBalanceMotion} is satisfied and that the
asymptotic behaviour far away from the interface of the structural
force balance \eqref{EQforceBalanceStructural} is satisfied. It
remains to be shown that the entire structural force balance profile
is satisfied across the interface. This implies (i) that there is no
further missing superadiabatic force contribution (within the current
approximations) and (ii) that there is no additional macroscopic force
exerted by the interface which could affect phase coexistence. In
other words, the interface is decoupled from the bulk.

Hence in order to proceed, we generalize the bulk expression
\eqref{EQnu3} for $\nu_3$ to inhomogeneous situations by choosing an
approximation that closely parallels the expression
\eqref{EQsphericalPressure} for $\Pi_2$, namely
\begin{align}
  \nu_3 &= \frac{\gamma }{2D_{\rm rot}}\Big[
  \frac{e_1}{\rho_{\rm jam}}v_{\rm loc}^2
  \rho_0  -
  \frac{e_2}{\rho_{\rm jam}^2}
  \nabla\cdot \frac{v_{\rm loc}^2}
              {(1-\rho_0/\rho_{\rm jam})^2}\nabla\rho_0
  \Big],
  \label{EQnu3ApproximationNEW}
\end{align}
where we have kept the bulk constant $e_1$ and we have introduced an
interfacial constant $e_2$. Furthermore $v_{\rm loc}$ is defined via
\eqref{EQvlocDefinition}, such that $\fv_{\rm sup,3}$ is a kinematic
functional. {\mm Here the second term on the right hand side of
  \eqref{EQnu3ApproximationNEW} is akin to the semi-local contribution
  in the van der Waals square gradient interfacial theory
  \cite{RowlinsonWidomBook}, generalized to possess a kinematic
  dependence on the flow via $v_{\rm loc}$.}

Figure~\ref{figPotentials}(b) presents results from theory and
simulation that illustrate the behaviour of both the adiabatic
chemical potential $\mu_{\rm ad}$, the ideal contribution $\mu_{\rm
  id}=k_BT\ln\eta'$, and the superadiabatic quiet life potential
$\nu_3$. We use the approximate equation of state
\eqref{EQadiabaticMu} in a local density treatment for $\mu_{\rm ad}$,
i.e.\ we replace $\rho_{\rm b}$ by $\rho_0$. Furthermore we use the
kinematic expression \eqref{EQnu3ApproximationNEW} with
\eqref{EQvlocDefinition} and \eqref{EQvfAsFunctional}.  We keep the
same value of the bulk parameter $e_1=0.0865$ as before and have set
the interfacial parameter {\ms $e_2=0.0385$.}  {\ms We can now search
  for the value of the interfacial width parameter $\lambda$ that
  leads to an optimal profile (as judged by minimial deviation from a
  constant {\mm value}). } The total chemical potential profile, as
the sum of ideal, adiabatic excess and quiet life contributions, is
{\ms indeed} constant to a very satisfactory degree, {\ms and the
  theoretical result for the interfacial profile matches the
  simulation data very well (cf.\ the blue solid and dashed lines in
  Fig.~\ref{figFourier}(a)).  That the theoretical result for the
  total chemical potential} deviates slighty from a constant value is
entirely consistent with the fact that the theoretical solution is
based on an ansatz for the density profile. The functional expressions
on the other hand constitute approximations, which we expect to lack
corrections as compared to the exact result. In summary, we have
demonstrated that the force balance equation is satisfied across the
interface.

\section{Conclusions}
In conclusion, we have developed a microscopic theory for bulk and
interfacial behaviour of active Brownian particles. The basis of our
treatment is the position- and orientation-resolved force density
balance. We have split the nonequilibrium contribution to the internal
force density into three contributions: (i) The drag force, which acts
in the opposite direction of the local flow direction. This is
strongly dependent on the local average density and possesses a
square-gradient correction, which models further drag due to motion in
an inhomogeneous density field. (ii) The intrinsic spherical pressure,
which acts in a similar way {\mm as} the equilibrium pressure in that
it provides additional repulsion, as generated from the internal
repulsive interactions.  The intrinsic spherical pressure has negative
values.  In the phase separated state, the dilute (dense) phase has
high (low) magnitude of the intrinsic spherical pressure. The internal
pressure is cancelled by the swim pressure that the polarized
interface exerts on the liquid. (iii) The quiet life internal force
field is of gradient form and it is independent of orientation. It
opposes the adiabatiatic force field, which arises solely due to the
density inhomogeneity and is defined via the adiabatic reference
system. The quiet life potential describes again additional
repulsion. Its magnitude comes from a moderate (linear) density
dependence and strong (quadratic) dependence on the local forward
speed. The prominent effect is that due to the fast motion in the
dilute phase, the quiet life potential is high.  In the slow dense
phase the quiet life potential is low. {\mz Hence the force field that
  emerges as the negative gradient of the quiet life chemical
  potential points towards the ``quiet'' liquid, as if the particles
  were aiming at a ``quiet life''.}  Although in both phases there
occurs additional repulsion, the net effect is a potential gradient,
which leads to a force acting from the gas into the liquid. For stable
phase-separated states, this force is balanced by the adiabatic
force. The balance constitutes a non-trivial condition, as the
adiabatic force solely depends on the density field (i.e.\ it is a
density functional) and the quiet life potential also depends on the
flow (i.e.\ it is a kinematic functional). As the flow is already
determined by (ii) and (iii) above, the non-trivial conditions
\eqref{EQcoexistenceMu} and \eqref{EQcoexistencePressure} for
stability of phase coexistence emerge. Technically, this can be
analysed with the standard tools of Maxwell construction.

{\ms The number of fit constants in our approach is low and comparable
  to what one needs in a square gradient theory of bulk and interface
  behaviour in equilibrium gas-liquid phase separation. Summarizing,
  we have used the definition of the effective packing fraction
  $\eta'$ (containing the number 0.8), the jamming density $\rho_{\rm
    jam}$, the strength of the quiet life chemical potential term
  $e_1$, cf.~\eqref{EQnu3} and \eqref{EQnu3ApproximationNEW} (where
  the same value of $e_1$ is used). Then the bulk forward speed
  $v_{\rm b}$, cf.~\eqref{EQvelocityLinearDecrease}, follows without
  further adjustable freedom. This makes three parameters for bulk
  coexistence (and the assumption of the scaled-particle equation of
  state in the description of the adiabatic reference system).
  In the interfacial treatment we introduce the parameter $\xi$ that
  determines the strength of the effect of spatial inhomogeneity on
  the average swim speed, and the strength $e_2$ of the interfacial
  contribution to the quiet life term,
  cf.~\eqref{EQnu3ApproximationNEW}.  Overall this makes $3+2=5$
  parameters for the microscopic description of both bulk and
  interface. There are no further hidden length, time, or energy
  scales. Any such dependence has been scaled out.
}

We {\mm successfully rationalized} all occurring bulk and interfacial
effects on the basis of a description which decouples the interfacial
contributions from the bulk coexistence conditions {\mz within the
  range of parameters considered}. Hence we conclude that {\mz within
  this range and within the gradient and power series approximations }
no coupling from interface back to the bulk {\it is required} in order
to describe the physics. This situation though does not rule out that
such a coupling exists \cite{solon2018rapComm,solon2018njp}. Our
theory should provide a convenient starting point for the
investigation of such interface-to-bulk coupling, as corresponding
physical effects can be incorporated. Besides the formal observations
of such effects, this would surely benefit from identifying {\it
  physical mechanisms} that would generate the coupling. Note that the
polarized interface alone does not necessitate any coupling. As we
have shown, the corresponding external pressure is balanced by the
internal spherical pressure, and both do not contribute to the
stability conditions. We leave the implications for general conditions
for phase coexistence \cite{krinninger2016prl} to future
work. Furthermore, taking full account of (small) ideal diffusion
contribution \cite{footnote2} to the dynamics is an interesting
problem, as is adding the description of shear viscous forces
\cite{prlVelocityGradient}, and relating to the concept of structural
force fields in more detail \cite{prlStructural}. {\mz Connections to
  work in driven lattice systems \cite{guioth2018}, in particular on
  phase coexistence far from equilibrium
  \cite{dickman2014,dickman2016}, and to the more general case of
  interacting dissipative units \cite{bertin2017topicalReview} are
  worth exploring.  It would also be interesting to investigate the
  effects of adding further external forces, such as those due to
  ramp-like external potentials considered in
  \cite{paliwal2018njp,guioth2019}. As the force balance without such
  a perturbation is already a delicate one, we expect profound changes
  upon such alterations, possibly similar to the changes that occur to
  equilibrium phase separation in confinement by external fields.

{\mm Further interesting connections to be made in future work include
  relating our approach to stochastic thermodynamics, as has been
  formulated for active particles by Speck \cite{speck2016epl} and to
  the interfacial findings of Bialk\'e et al.~\cite{bialke2015}; work
  along the latter lines is in progress \cite{hermann2019tension}.
  Furthermore investigating within our theory the relationship to the
  Gibbs-Thomson relation, as considered by Lee \cite{lee2017}, could
  be worthwhile, as would be to consider curvature-dependence, as
  performed by Patch et al.~\cite{patch2018}, and depletion forces in
  nonequilibrium \cite{dzubiella2003}.
 {\mc A finite-size analysis in the present paper has to remain open.}
  It would be interesting to study in future simulation work the
  finite-size dependence of the superadiabatic force fields. For a
  study of finite-size effects in the critical region see
  \cite{binder2018}; {\mb for an investigation of finite-size effects
    on the pressure see \cite{patch2017}}. The correlator expression
  developed in this work could provide the backbone of such work.}

}

\acknowledgments We thank Marjolein Dijkstra, Ren\'e van Roij, Jeroen
Rodenburg, Siddharth Paliwal, and Martin Oettel for useful discussions
and Bob Evans for useful discussions and for having been instrumental in
developing the terminology.

\appendix
{\ms
\section{One-body equation of motion}
\label{APPoneBodyFromSmoluchowski}
We derive the force density balance \eqref{EQforceDensityBalance} from
the underlying Fokker-Planck equation of motion for the many-body
probability distribution function $\Psi(\rv^N,\ov^N,t)$, where
$\rv^N\equiv\rv_1,\ldots,\rv_N$ denotes the set of all position
coordinates and $\ov^N\equiv\ov_1,\ldots,\ov_N$ denotes the set of all
particle orientations; the corresponding momenta are irrelevant
degrees of freedom due to the overdamped nature of the dynamics. This
(Smoluchowski) equation of motion is analogous to the Langevin picture
\eqref{EQlangevinPositions} and \eqref{EQlangevinOrientations}, and
given by a many-body continuity equation of the form
\begin{align}
  \frac{\partial\Psi}{\partial t} &= -\sum_i \nabla_i\cdot\vel_i \Psi
  -\sum_i \nabla_i^\omega\cdot\vel_i^\omega \Psi,
  \label{EQsmoluchoswkiABP}
\end{align}
where the sums run over all particles. Here the translational
configurational velocity $\vel_i(\rv^N,\ov^N,t)$ and the rotational
configurational velocity $\vel_i^\omega(\rv^N,\ov^N,t)$ of
particle~$i$ are many-body functions given, respectively, by
\begin{align}
  \gamma\vel_i &= -\nabla_i u + \gamma s \ov_i
  -k_BT \nabla_i \ln \Psi,\\
  \gamma^\omega \vel_i^\omega &=
  -k_BT \nabla_i^\omega \ln \Psi,
\end{align}
where $u(\rv^N)=\sum_j\sum_{k(\neq j)}\phi(|\rv_j-\rv_k|)/2$ is the
total internal potential energy, and $\nabla_i^\omega$ denotes the
derivative with respect to orientation $\ov_i$.

The microscopic definitions for the one-body distribution functions
are as follows. The density distribution is
$\rho(\rv,\ov,t)=\langle\sum_i\delta_i\rangle$, where
$\delta_i=\delta(\rv-\rv_i)\delta(\ov-\ov_i)$ with the Dirac
distribution $\delta(\cdot)$, and the angles denote a statistical
average, which in the Smoluchowski picture is defined as
$\langle\cdot\rangle=\int d\rv^N d\ov^N \cdot \Psi(\rv^N,\ov^N,t)$.
The one-body current is $\Jv(\rv,\ov,t)=\langle\sum_i
\delta_i\vel_i\rangle$, where $\vel_i$ is the translational velocity
of particle $i$ at time $t$. The internal one-body force density field
is $\Fv_{\rm int}(\rv,\ov,t)=-\langle\sum_i\delta_i\nabla_i u\rangle$.

In order to obtain the one-body dynamics, we differentiate in time the
(definition of the) one-body density distribution,
\begin{align}
  \frac{\partial \rho}{\partial t} &=
  \int d\rv^N d\ov^N \sum_i \delta_i 
  \frac{\partial \Psi}{\partial t}.
  \label{EQdensityTimeDerivative}
\end{align}
Next we replace the time derivative of the many-body distribution
function with the right hand side of the Smoluchowski equation
\eqref{EQsmoluchoswkiABP} and integrate by parts in both positions and
orientations. By using the identities
$\nabla_i\delta_i=-\nabla\delta_i$ and $\nabla_i^\omega
\delta_i=-\nabla^\omega \delta_i$, it is straightforward to rewrite
\eqref{EQdensityTimeDerivative} in the form of the continuity equation
\eqref{EQcontinuity} with the translational current given by
\eqref{EQforceDensityBalance} and the rotational current being that of
free rotational diffusion. 

Some more details, also about power functional theory for active
Brownian particles, and more generally orientation-dependent models
can be found in \cite{krinninger2019jcp}; an overview of different
methods to sample the one-body current in BD simulations is given in
\cite{delasheras2018customFlow}.}

{\ms
\section{Classical density functional theory}
\label{SECdft}
In a one-component equilibrium system of spheres, according to
classical density functional theory \cite{evans1979}, the equilibrium
one-body density distribution $\rho(\rv)$ is obtained from the
solution of
\begin{align}
  k_BT \ln \rho(\rv)
  + \frac{\delta F_{\rm exc}[\rho]}{\delta \rho(\rv)}
  &= \mu - V_{\rm ext}(\rv).
  \label{EQappendixBEulerLangrange}
\end{align}
Here the irrelevant thermal de Broglie wavelength has been set to
unity.  Eq.~\eqref{EQappendixBEulerLangrange} represents a
self-consistency relation for the density profile $\rho(\rv)$. The
equation results from the minimization principle for the grand
potential functional $\Omega[\rho]$, which states that $\Omega$ has
its minimal value at the physical equilibrium density. Here the
functional maps the position-dependent function $\rho(\rv)$ onto the
number $\Omega$. The grand potential functional is given as a sum of
intrinsic and external contributions, according to
\begin{align}
  \Omega[\rho] &= k_BT \int d\rv \rho(\rv)(\ln \rho(\rv)-1)
  + F_{\rm exc}[\rho]\notag\\&\qquad
  +\int d\rv \rho(\rv)(V_{\rm ext}(\rv)-\mu).
  \label{EQappendixBOmega}
\end{align}
Here the first term on the right hand side is the Helmholtz free
energy functional of the ideal gas, the second term $F_{\rm
  exc}[\rho]$ is the excess (over ideal gas) intrinsic contribution
due to the interparticle interactions $u(\rv^N)$ and the third term
represents the external potential energy and includes the chemical
potential contribution. This framework is formally exact and both
$\Omega[\rho]$ and $F_{\rm exc}[\rho]$ have a microscopic definition
\cite{evans1979} that renders them uniquely defined mathematical
objects. Equation~\eqref{EQappendixBEulerLangrange} follows from
\eqref{EQappendixBOmega} by the condition of vanishing first
derivative, i.e.\ calculating the functional derivative $\delta
\Omega/\delta \rho(\rv)=0$, as is appropriate at the minimum. In the
present study we use this framework to describe the adiabatic
reference state. Hence in our application we set
$\rho(\rv)=2\pi\rho_0(\rv)$, where $\rho_0(\rv)$ is the angular
average of the orientation resolved density distribution of the active
Brownian particles.
}

{\mb
\section{Simulation box geometry}

In order to illustrate the dependence of the simulation results on the
simulation box geometry, we show in Fig.~\ref{FIGBoxVariation}(a) the
isotropic component of the density profile, $\rho_0(x)$, for different
values of the average density $N\sigma^2/V$ with $s\tau/\sigma = 60$,
$k_BT/\epsilon = 0.5$ and the aspect ratio ${\cal A}$ of the length of
the simulation box in the $x$- and in the $y$-directions, ${\cal A} =
5$ being fixed. For overall density $N\sigma^2/V = 0.5$ the system
does not separate into two phases and $\rho_0(x) = {\rm const} =
N\sigma^2/V$. There is a very small increase in local density near the
center of the simulation box, which is an artifact introduced by
fixing the center of mass of the entire system, which is a means to
stabilize the interface position(s).  Increasing the average density
leads to phase separation. The coexisting densities are rather
independent of the value of the bulk density, but the relative
fraction of the dense phase increases upon increasing the overall
density. Note that the total volume of the simulation box decreases
with increasing the average density $N\sigma^2/V$, as we keep $N$
fixed.

In Fig.~\ref{FIGBoxVariation}(b) we display the dependence of $\rho_0$
on the simulation box aspect ratio ${\cal A}$. The other parameters
are kept fixed: $k_BT/\epsilon = 0.5$, $s\tau/\sigma = 60$, and
$N\sigma^2/V = 0.7$. For ${\cal A} = 2.5$ and ${\cal A} = 5$ the
density profiles share the same overall shape and the coexisting bulk
densities in the gas and in the liquid are the same. However, for
${\cal A}=2.5$ the liquid slab is already very thin and the two
interfaces become very close to each other and not as well decoupled
from each other as in the case ${\cal A}=5$. Increasing the aspect
ratio further to ${\cal A} = 10$, i.e.\ making the simulation box
narrower, the shape of $\rho_0$ is not fully retained and a minimum
develops at the center of the simulation box. We assume that finite
size effects due to the short length of the simulation box in the
$y-$direction are responsible for this artifact. Inspection of
snapshots reveals that typical configurations also involve the
nucleation of an additional gas region at the center of the box. Hence
a (periodic) succession of gas-liquid-gas-liquid-gas regions
appears. For such states again the localization of the interface
fails. Nevertheless, the plateau values of the density profile suggest
that the bulk densities in the gas and in the liquid phases are
similar for all aspect ratios considered.

\label{SECboxGeometry}
\begin{figure}
  \includegraphics[width=0.48\textwidth]{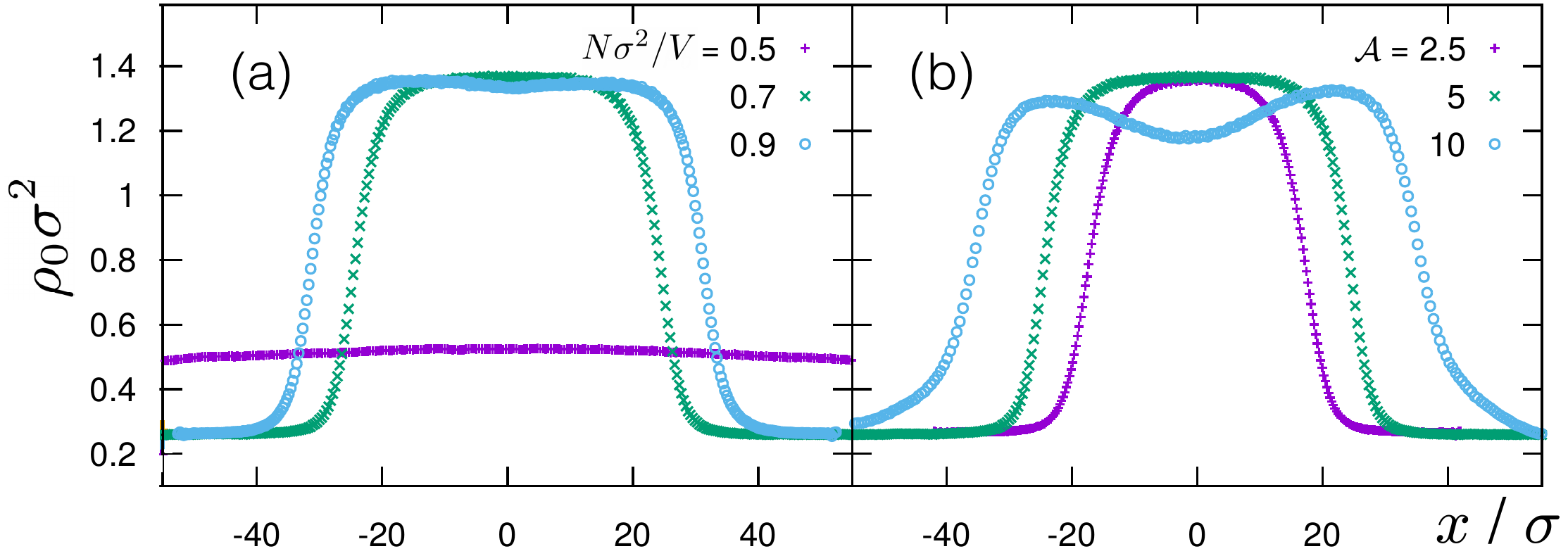}
  \caption{{\mb Isotropic Fourier component of the scaled density
      profile, $\rho_0\sigma^2$, as a function of $x/\sigma$ obtained
      from BD simulations for $s\tau/\sigma=60$ and $k_BT/\epsilon =
      0.5$.  (a) For different values of the average density
      $N\sigma^2/V= 0.5, 0.7$, and 0.9 (as indicated) with simulation
      box aspect ratio ${\cal A} = 5$. (b) For different aspect ratios
      ${\cal A} = 2.5, 5,$ and 10, with average density $N\sigma^2/V =
      0.7$.}}
\label{FIGBoxVariation}
\end{figure}

\section{Square gradient strength $\xi$}

We display in Fig.~\ref{FIGxiVariation} the Fourier coefficients of
the density profile, $\rho_n(x)$, for two further values of the
strength of the square density gradient term: $\xi=0$ (a), which is
identical to omitting the square gradient term in
Eq.~\eqref{EQvfAsFunctional}, and as a further representative case
$\xi=2100$ (b). As a reference the results for our (optimal) parameter
choice $\xi=700$ are also shown (black solid lines); these data are
identical to that shown in Fig.~\ref{figFourier}(a).

It is clear that very large values of $\xi$ introduce artifacts, such
as e.g.\ the double hump in the polarization profile $\rho_1$. For
very small values of $\xi$ (with zero being an extreme case thereof),
the overall amplitudes become exaggerated. The chosen value $\xi=700$
represents a compromise where neither of the two effects is dominant.

\label{SECsquareGradientParameter}
\begin{figure}
  \includegraphics[width=0.48\textwidth]{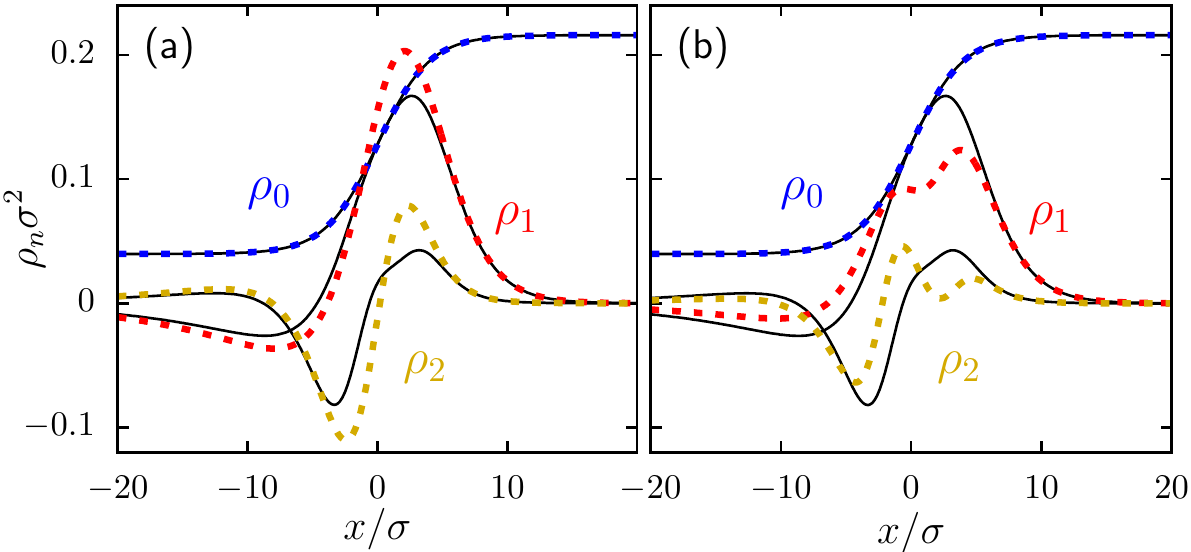}
  \caption{{\mb Fourier coefficients $\rho_n\sigma^2$ of the density
      distribution as a function of $x/\sigma$ across the
      interface. Shown are the theoretical results for $n=0,1,2$
      (dashed lines, as indicated) for different values of the
      parameter $\xi=0$ (a) and 2100 (b). As a reference the results
      for $\xi=700$ are also shown (solid black lines); these results
      are identical to the data in the right panel of
      Fig.~\ref{figFourier}(a). }}
\label{FIGxiVariation}
\end{figure}

}


\begin{thebibliography}{31}
\bibitem{footnote2} Going beyond this approximation requires to
  introduce two further superadiabatic force density fields, $\Fv_{\rm
    sup,4}$ and $\Fv_{\rm sup,5}$. These contribute to the total
  internal force density, such that $\Fv_{\rm int}=\Fv_{\rm
    ad}+\sum_{i=0}^5 \Fv_{{\rm sup},i}$. The additional fields are
  kinematic functionals and are specfified by the cancellations
  $\Fv_{\rm sup,4}=k_BT\nabla\rho$ and $\Fv_{\rm sup,5}=-k_BT
  (\rho/\rho_0) \nabla \rho_0$. Then in the correlator expressions
  \eqref{EQFsup0PrimeAsFunctional}, \eqref{EQFsup1PrimeAsAverage},
  \eqref{EQFsup2AsCorrelator} and \eqref{EQFswimAsCorrelator} the full
  internal force density field $\Fv_{\rm int}$ needs to be replaced by
  $\Fv_{\rm int}'=\Fv_{\rm int}-\Fv_{\rm sup,4}$.
\bibitem{footnote4} The density distribution in a standard DFT of
  spheres with no orientations is given by $2\pi\rho_0$.
\bibitem{footnote3} The polynomial is $ -15 c^2 \eta_{\rm c}^4 + 4(4
  c^2-3 c)\eta_{\rm c}^3 + (-3c^2+16c+9)\eta_{\rm c}^2 + 4(c+2)
  \eta_{\rm c} +1 =0, $ where $c=0.8$ and $\eta_{\rm c}$ indicates the
  critical packing fraction.
\bibitem{footnote5} This form implies that the $x$-axis points towards
  the dense phase, as is the case in the present setup.
\end{thebibliography}
\end{document}